\documentclass[conference]{IEEEtran}
\usepackage[colorlinks=true,linkcolor=blue,citecolor=purple]{hyperref}
\usepackage{array}
\usepackage{tcolorbox}

\usepackage[square,sort&compress,comma,numbers]{natbib}
\usepackage{threeparttable, multirow, color, algorithmic, algorithm, url, balance, graphicx, booktabs, amsfonts, balance, color, amssymb }
\usepackage{makecell}
\usepackage[misc,geometry]{ifsym}

\def\BibTeX{{\rm B\kern-.05em{\sc i\kern-.025em b}\kern-.08em
    T\kern-.1667em\lower.7ex\hbox{E}\kern-.125emX}}
    
\graphicspath{{pics/}}

\makeatletter
\def\thanks#1{\protected@xdef\@thanks{\@thanks
        \protect\footnotetext{#1}}}
\makeatother

\begin{document}
%
\title{UltraFuzz: Towards Resource-saving in Distributed Fuzzing }


\author {\IEEEauthorblockN{Xu Zhou, Pengfei Wang \Letter , Chenyifan Liu, Tai Yue, Yingying Liu, Congxi Song, Kai Lu, Qidi Yin, Xu Han}
\thanks {Pengfei Wang (pfwang@nudt.edu.cn) is the corresponding author.}
\IEEEauthorblockA{National University of Defense Technology}}





%


\maketitle

\begin{abstract}

Recent research has sought to improve fuzzing performance via parallel computing.
However, researchers focus on improving efficiency while ignoring the increasing cost of testing resources. 
Parallel fuzzing in the distributed environment amplifies the resource-wasting problem caused by the random nature of fuzzing. In the parallel mode, owing to the lack of an appropriate task dispatching scheme and timely fuzzing status synchronization among different fuzzing instances, task conflicts and workload imbalance occur, making the resource-wasting problem severe.
In this paper, we design UltraFuzz, a fuzzer for resource-saving in distributed fuzzing. Based on centralized dynamic scheduling, UltraFuzz can dispatch tasks and schedule power globally and reasonably to avoid resource-wasting.
Besides, UltraFuzz can elastically allocate computing power for fuzzing and seed evaluation, thereby avoiding the potential bottleneck of seed evaluation that blocks the fuzzing process.
UltraFuzz was evaluated using real-world programs, and the results show that with the same testing resource, UltraFuzz outperforms state-of-the-art tools, such as AFL, AFL-P, PAFL, and EnFuzz.
Most importantly, the experiment reveals certain results that seem counter-intuitive, namely that parallel fuzzing can achieve ``super-linear acceleration'' when compared with single-core fuzzing. 
We conduct additional experiments to reveal the deep reasons behind this phenomenon and dig deep into the inherent advantages of parallel fuzzing over serial fuzzing, including the global optimization of seed energy scheduling and the escape of local optimal seed. Additionally, 24 real-world vulnerabilities were discovered using UltraFuzz.

\end{abstract}


%
\IEEEpeerreviewmaketitle

\section{Introduction}

Software vulnerabilities are significant threats to information systems. 
Security analysts leverage a huge amount of resources at discovering software
vulnerabilities, a process that is usually resource-hungry, time-consuming, and labor-intensive. Among the various techniques available, fuzzing is effective and has been widely used to test multi-threaded programs \cite{zhaokrace, vinesh2020confuzz, jeong2019razzer}, libraries \cite{blair2020hotfuzz}, kernel \cite{song2019periscope,kimhfl}, protocols \cite{yu2019poster, zheng2019firm}, and smart contracts \cite{jiang2018contractfuzzer, wustholz2019targeted}.
Fuzzing usually generates massive random \emph{test cases} to run the target program and monitors running crashes to report vulnerabilities. To thoroughly test a program, countless test cases are generated and the program is executed repeatedly~\cite{fuzzingStateOfTheArt,fuzzingArt}. However, due to the randomness of fuzzing, large numbers of redundant test cases exercise the same path, wasting testing resources. As a result, this process is very time consuming, and usually costs many hours of computation, even days or months. 

Therefore, fuzzing requires noticeable efficiency improvement to make the process of vulnerability detection more timely.
Most research thus far has sought to improve fuzzing performance by designing novel algorithms. These algorithms improve performance by optimizing the core mechanism of fuzzing, including seed generation~\cite{wang2016seededfuzz,wang2020tofu,jain2018tiff,you2019profuzzer}, mutation strategy~\cite{chen2018hawkeye,li2019v,you2017semfuzz,you2019profuzzer}, seed prioritization~\cite{yue2020ecofuzz,bohme2017directed,wang2020tofu}, etc. However, the extent to which algorithms can improve efficiency is limited. On average, prevalent AFL-based fuzzers, such as Fairfuzz~\cite{fairfuzz}, AFLFast.new~\cite{AFLFastnew}, and FidgetyAFL~\cite{fidgetyAFL}), only increase efficiency by around 15\% when compared with original AFL. 
Another research direction to increase fuzzing efficiency is utilizing parallel computing resources to process fuzzing workloads concurrently. Since fuzzing workloads do not have much data dependency, we can foresee a great efficiency increase during parallel fuzzing. In an ideal situation, performance could be improved 100\% simply by doubling computing capacity. However, most researchers focus on improving efficiency to save time but ignore the increasing cost of testing resources. Take Google's OSS-Fuzz project as an example, it leverages more than 25,000 machines that process an average of ten trillion test inputs a day. As a result, it has found 16,000 bugs in Chrome and 11,000 bugs in over 160 projects in two years~\cite{bohmefuzzing}. Though effective, such huge-scale fuzzing is extremely resource-consuming. Averagely, to discover a bug, OSS-Fuzz needs 1.85 machines running for one year.
This is because real-world parallel fuzzing is not simply a pile-up of computing resources, and parallel fuzzing in a large-scale distributed environment can amplify the resource-consuming problem. 


We summarize three situations that can cause resource-wasting in distributed fuzzing.
(1) Owing to the absence of global scheduling and timely update of fuzzing status (e.g., seeds and coverage) among the working nodes, different fuzzing instances may mutate the same seed and generate redundant test cases that execute the same path, causing task conflicts and wasting testing resources. This is the root cause of resource wasting. (2) To reduce task conflicts, new seeds generated by different fuzzing instances should be evaluated and de-duplicated. However, such a seed evaluation process can become a bottleneck when huge numbers of new seeds are generated dramatically by multiple fuzzing instances, and the inappropriate computing power allocation between fuzz testing and seed evaluation can either restrict the fuzzing process or waste the resource. (3) When each computing core has unequal computing capability or the number of computing cores dynamically changes, inflexible task scheduling can cause workload imbalance, which also leads to resource-wasting. Moreover, in 2020, B\"ohme \textit{et al.}~\cite{bohmefuzzing} proposed an empirical law in fuzzing, which was that
``\textit{given the same non-deterministic fuzzer, finding linearly more bugs in the same time requires exponentially more machines}.'' This law implies that the huge consumption of computing resources has restricted the scalability of distributed fuzzing like a wall, which demands a prompt solution.

In this paper, we focus on the resource-wasting problem in distributed fuzzing and optimize the fuzzing process towards not only efficiency but also resource-saving. To solve the task conflict problem caused by redundant test cases, we propose to use a centralized dynamic scheduling scheme to evaluate seeds and dispatch tasks in a centralized manner. We separate scheduling from fuzzing and collect the new seeds from all the fuzzing instances to filter out duplicates and prioritize them based on their evaluation. 
We regard parallel fuzzing as the global optimization of seed selection, energy scheduling, fuzzing status synchronization, and task dispatching among all the working nodes. 
To improve synchronization efficiency, we use a hierarchical scheme based on fuzzing information characteristics, which achieves instant synchronization.
For the seed evaluation requirement caused by the dramatic increase of new seeds, we classify the roles of working nodes into \textit{evaluating instances} and \textit{fuzzing instances}. Evaluating instances filter out redundant test cases, while the fuzzing instances run the tests. An elastic computing power allocation scheme is adopted to handle the dramatic increase of seed evaluation requests, which can adaptively coordinate the two groups of instances and achieves a global optimization of resource allocation. Finally, for the workload imbalance caused by dynamic computing power change, we dispatch fuzzing tasks with a dynamic on-demand scheme, which can adaptively fit the computing capability of fuzzing instances and is compatible with environmental change.
In summary, we make the following contributions:

\textbf{We conduct the first research on resource-saving in distributed fuzzing.} Though improves efficiency, distributed environment amplifies the resource-wasting problem of fuzzing. We summarize three situations in distributed fuzzing that can cause resource-wasting and propose novel approaches to solve these problems. Based on a dynamic centralized task scheduling scheme, our approach can achieve instant fuzzing status synchronization, global energy scheduling, and elastic computing resource allocation, which can reduce resource wasting by avoiding problems such as task conflicts, workload imbalance, and potential bottlenecks of seed evaluation.


\textbf{We implement our design and evaluate it in large-scale experiments.} 
We implement a tool called UltraFuzz and conduct experiments on 16 real-world programs with computing cores ranging from 8 to 128. The scale of our experiments is more than 6,186  CPU days. Results show that UltraFuzz outperforms state-of-the-art tools like AFL, AFL-P, PAFL, and EnFuzz in aspects of branch coverage, path coverage, and the number of test cases. The synchronization overhead is about 1.26\%. UltraFuzz also found 24 vulnerabilities in them. 
UltraFuzz has been made publicly available online \footnote{\url{https://gitlink.org.cn/hunter-2018/Ultrafuzz.git}} to encourage and support future research. 

 \textbf{We observe counter-intuitive ``super-linear acceleration''.} 
Intuitively speaking, parallel fuzzing should only achieve sub-linear acceleration. For example, for parallel fuzzing with four nodes in one hour and fuzzing with one node in four hours, the former should perform less well, owing to the cost of synchronization and task dispatching. However, our experiments show a different result. We conduct additional experiments to reveal the deep reasons behind this phenomenon and dig deep into the inherent advantages of parallel fuzzing over serial fuzzing, including the global optimization of seed energy scheduling and the escape of local optimal seed.



\section{Background}

\subsection{American Fuzzy Lop}

AFL (American fuzzy lop) \cite{AFL1} is a widely used coverage-based greybox fuzzer, and many state-of-the-art greybox fuzzers \cite{bohme2017coverage, lyu2019mopt, yue2020ecofuzz, yue2019learnafl} are built on top of it. AFL uses lightweight instrumentation to capture basic block transitions and gain coverage information during run-time. Compared to other instrumented fuzzers, AFL has a modest performance overhead, and we determined to base our design on AFL.


AFL leverages the edge-coverage information to select seeds. It maintains a seed queue and mutates the seed to generate testcases. If a testcase exercises a new path, it is added to the queue as a new seed.  In addition to covering a new branch, when a seed is smaller and executes quicker with respect to previous test cases that hit a given branch, it is also labeled as \textit{favored} and prioritized. 
AFL uses a bitmap with edges as keys and top-rated seeds as values to maintain the best performance seeds for each edge. It selects favored seeds from the top-rated queue and gives them more fuzzing chances (i.e., energy) than the non-favored ones.
AFL assigns energy to the seeds according to the performance score of each seed, which is based on coverage (prioritizing inputs that cover new paths), execution time (prioritizing inputs that execute faster), and discovery time (prioritizing inputs discovered later). In particular, if a test case exercises a new path, AFL will double the energy assigned to it. Fuzzing tasks are handled by AFL in the form of a selected seed and its assigned energy.

The seed selection strategy and energy schedule algorithm play important roles in improving the efficiency of fuzzing. Some works \cite{bohme2017coverage, yue2020ecofuzz} have proved that selecting the seeds exercising rare paths and focusing on them can generate more test cases to trigger new states than focusing on seeds exercising high-frequency paths. They also pointed out that an efficient schedule algorithm is based on the evaluation of each seed, which is determined by the information of coverage and seeds (e.g., the execution speed). However, in distributed fuzzing, the way to optimize the scheduling algorithm is different from single-core running. Compared to fuzzing with a single core, such information can't be shared timely in distributed fuzzing. Thus, it is crucial to building an efficient mechanism of information synchronization to support the global seed selection and energy schedule in distributed fuzzing.

\subsection{Parallel Fuzzing}


Parallel fuzzing first evolves from a naive method, i.e., the parallel mode of AFL (i.e., AFL-P), which simply runs multiple fuzzer instances with the same target simultaneously. Each instance of AFL binds a core and periodically re-scans the top-level sync directory for any test cases found in other instances. \emph{Multi-core parallel fuzzing} represents the evolution of the naive method. However, the parallelism at this stage is limited by intra-machine. Distributed fuzzing~\cite{roving,disafl} extends intra-machine parallelism to multiple machines connected by a network, allowing using more computing resources. Then, the researcher takes a step further by sharing seeds between fuzzer instances.
In this mode, the program starts to schedule tasks between fuzzing instances to alleviate task conflicts \cite{pfuzz1, PAFL}. 
To optimize parallel fuzzing towards resource-saving, we still need to overcome the following technical challenges.

(1) \textbf{Task conflicts}. In parallel fuzzing, owing to the absence of global scheduling and timely update of fuzzing information among the instances, different fuzzing instances may mutate the same seeds or generate redundant test cases that execute the same path, causing task conflicts, which will lead to a severe waste of testing resources. Though some tools, such as P-fuzz~\cite{pfuzz1} and PAFL~\cite{PAFL}, try to synchronize fuzzing information among fuzzing instances, a centralized scheduling scheme is still required to optimize parallel fuzzing from a global perspective.

(2) \textbf{Synchronization overhead}. In parallel fuzzing, various fuzzing information can be shared among fuzzing instances to increase performance, such as seeds, coverage, hangs, and crashes. However, the synchronization overhead inevitably deducts the performance, and such deduction becomes more severe as instances increase. Thus, what information to share and how to share it is critical, especially in a distributed environment.
To alleviate performance deduction and improve resource efficiency, some approaches~\cite{roving,disafl,enfuzz,pfuzz1} adopt periodical synchronization to achieve a tradeoff between effectiveness and efficiency. However, timely updating of fuzzing status is critical to avoid task conflict. 


(3) \textbf{Computing power allocation}. When dispatching fuzzing tasks to different fuzzing instances, computing power allocation is another challenge. 
For example, AFL-P and PAFL~\cite{PAFL} are only designed for intra-machine multi-core parallel fuzzing, which assumes the computing capability of all cores is equal and assigns equal workloads to each core. However, for a distributed fuzzing environment, each computing core might have a different computing capability and the number of computing cores might dynamically change, such a static strategy might aggravate workload imbalance and waste resources. Besides, with the increasing of fuzzing instances, large bunches of new seeds can over-burden the system, and the seed evaluation (de-duplication) becomes a bottleneck that needs a dynamical re-allocation of the computing power.

In this paper, we propose to solve the task conflict problem via a centralized dynamic scheduling scheme. By separating scheduling from fuzzing, centralized dynamic scheduling can select seeds and schedule power from a global perspective. To improve synchronization efficiency, we use a hierarchical scheme based on fuzzing information characteristics, which achieves instant synchronization. For the workload imbalance caused by dynamic computing power change, we dispatch fuzzing tasks with a dynamic on-demand scheme. Finally, an elastic computing power allocation scheme is adopted to handle the dramatic increase of seed evaluation requests, which can adaptively coordinate the two groups of instances and achieve global optimization of resource allocation.

\section{Design}


\subsection{Centralized Dynamic Scheduling}

\begin{figure}[h]
  \centering
  \includegraphics[width=0.9\linewidth]{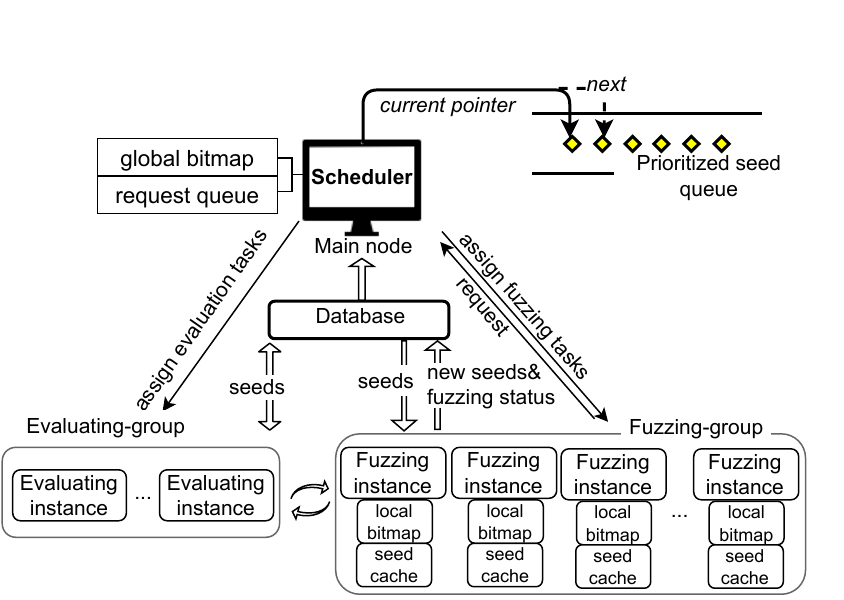}
  \caption{Overview of centralized dynamic scheduling.}
  \label{design}
\end{figure}

We propose to use centralized dynamic scheduling to optimize distributed fuzzing globally from aspects of seed selection, energy scheduling, fuzzing status synchronization, and task dispatching. 
As Fig.~\ref{design} shows, the fuzzing architecture consists of four components: a \emph{scheduler}, a \emph{database}, the \emph{fuzzing instances}, and the \emph{evaluating instances}. Among the computing nodes that make up the distributed system, one is selected as the \emph{main node}, and the rest are \emph{working nodes} that are used to conduct fuzzing or to evaluate the seeds.

On the main node, a scheduler is designed for scheduling seed evaluation, prioritizing the seeds in a queue, processing requests from the fuzzing instances, and dispatching fuzzing tasks. A fuzzing task contains two kinds of information: the index of the seed (i.e. the hash value) and an integer indicating how many times the seed should mutate (i.e. the energy).  Meanwhile, a database is deployed on the main node to store and share fuzzing data (e.g., seeds, fuzzing status, and coverage). Each fuzzing instance connects to the database to synchronize fuzzing information and seeds. The synchronization scheme will be introduced in Section \ref{sync}.

For a working node, it can either works as a fuzzing instance or an evaluating instance. A fuzzing instance is responsible for running test cases and mutating seeds. It downloads task seeds assigned by the scheduler from the database and uploads new seeds to the database. The majority of computing cores are used to run fuzzing instances. An evaluating instance works to filter out the duplicate seeds and the fuzzing instances download unique ones. 
Both the evaluating instances and the fuzzing instances are connected to the scheduler and the database.  
The scheduler dispatches evaluating tasks to evaluating instances and fuzzing tasks to fuzzing instances. Given the fact that the number of new seeds can change dramatically, the role of evaluating instances and fuzzing instances can dynamically switch based on the requirements of the evaluating tasks, which alleviates potential over-burdening of seed evaluation and achieve the best performance possible. This scheme will be introduced in Section \ref{dynamic}.

From the perspective of the scheduler, each time a fuzzing instance discovers a new seed, it will be stored in the database and evaluated by the evaluating instance. Then, the scheduler sorts seeds in a seed queue based on the discovered time and dispatch fuzzing tasks according to their fuzzing status (i.e., how important a seed is and how many times it has been fuzzed).  
Since this work concentrates on optimizing the architecture of distributed fuzzing, we carry over the original seed selection scheme of AFL. Namely, we favor seeds that are small in size and execute fast. When traversing the seed queue, such favored seeds will have a higher probability to be selected for further mutation. Similarly, we inherit the power scheduling of AFL, the mutation chances are determined by the product of some parameters in AFL, including \texttt{execution\_time}, \texttt{bitmap\_size}, \texttt{handicap}, and \texttt{depth}. 

This centralized dynamic scheduling separates scheduling from fuzzing, which can select seeds and schedule power from a global perspective, alleviating the problem of resource-wasting caused by task conflicts, and extending intra-machine parallel fuzzing to inter-machine in a distributed environment.

\subsection{On-demand Task Dispatching}




In our design, fuzzing tasks are dispatched via a dynamic on-demand scheme. Each fuzzing instance makes a request to the scheduler for a fuzzing task as soon as it is free. The scheduler stores the requests in a queue based on the request time. Then it selects the seed with the highest priority and responds to the front fuzzing request in the request queue with a task specification.
Only when a fuzzing task is requested, will the seed be selected and the power is calculated and assigned by the scheduler.
A fuzzing instance will request a new task when the current task is about to finish, namely when the last execution of the current seed begins. Once a new task is assigned, the fuzz instance continues to finish the current task and then switches to the new task. Thus, for some cases, the waiting time between task-dispatching can be saved, which is more efficient.
Under this scheme, workloads are balanced automatically and important tasks would be done first. Such dynamic task dispatching is flexible and compatible with testing environment change. For example, if we increase or decrease the test machines, the dynamic task dispatching scheme can adjust to the new machines automatically and dispatch tasks according to the new machine number.

\begin{figure}[h]
  \centering
  \includegraphics[width=0.9\linewidth]{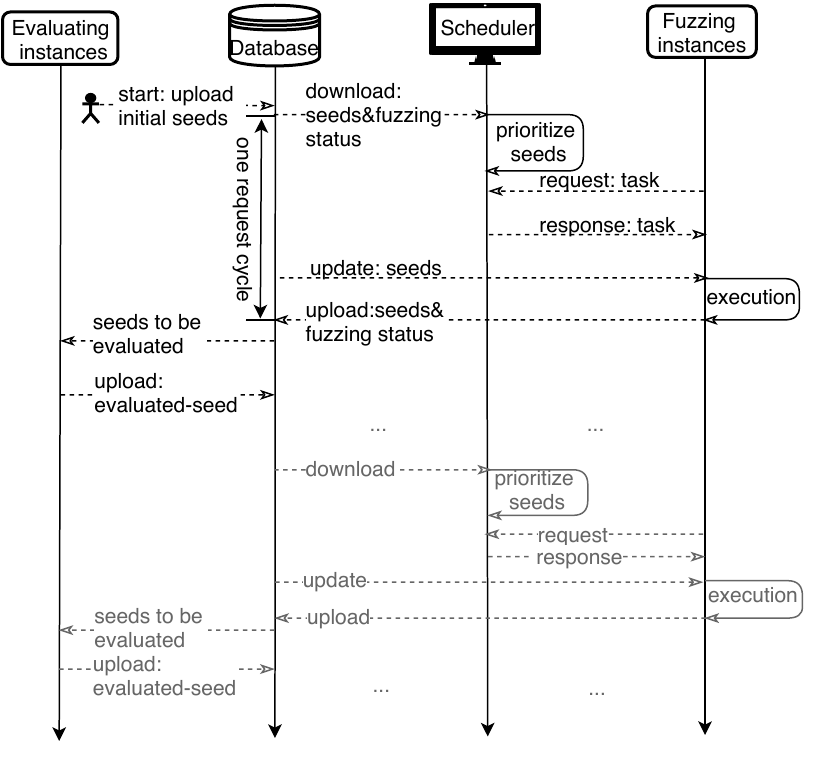}
  \caption{Procedure of centralized dynamic scheduling.}
  \label{procedure}
\end{figure}

The task dispatching is handled by the scheduler in the main node. 
As Fig. \ref{procedure} shows, the whole process starts with the scheduler downloading seeds from the database and prioritizing them in a seed queue. Then the idle fuzzing instances request fuzzing tasks from the scheduler, and the scheduler responds to the sorted request with a task specification, including the seed index and the number of mutations (i.e., energy). With the seed index, the fuzzing instance downloads the seed from the database and conducts the fuzzing tests. 
When a test reveals a new seed, the fuzzing instance will upload it to the database. 
Meanwhile, the scheduler arranges evaluating instances according to the number of seeds to be evaluated. Evaluating instances download seeds from the database to filter out duplicates and remove them from the database.

\subsection{Instant Hierarchical Synchronization}
\label{sync}
In parallel fuzzing, fuzzing status synchronization among different fuzzing instances has always been a challenge. Efficient fuzzing status synchronization
can reduce runtime overhead and assist global scheduling to save computing resources. Our design uses a database instead of a file system to synchronize fuzzing information. This is because file-based synchronization does not scale well in an inter-machine mode~\cite{Wen2017Designing}. 
We mainly synchronize three kinds of fuzzing information: \textit{seeds}, \textit{fuzzing status}, and \textit{coverage information}. We propose a hierarchical fuzzing status synchronization scheme according to the characteristics of each information type, whilst also ensuring performance is not compromised. 

\textbf{Fuzzing status direct sharing}.
Fuzzing status determines the priority of a seed and how many times it should be fuzzed. It is evaluated mainly by the number of times it has been fuzzed, including the \emph{depth} - the generation of the seed from the initial seed, \emph{handicap} - fuzzing cycles the queue has been done, and \emph{bitmap\_size} - the number of seeds' bits used for mutation. 
Although fuzzing status is stored in the database alongside the seeds, it is much different from seeds. The size of a seed is usually several KB or more, while the size of a fuzzing status is much smaller. Seeds are used by fuzzing instances and can be updated incrementally, while only the scheduler uses fuzzing status. While seeds are constant once they are written in the database, fuzzing status is updated each time the seed is fuzzed, which is much more frequent. For data as lightweight and frequently used as this, it is practical to share directly through the database. Thus we separate the use of fuzzing status from the corresponding seeds. The scheduler dispatches tasks based on fuzzing status instead of real seeds. It only maintains a lightweight queue to prioritize the seeds, which will greatly reduce the scheduler's network pressure when dispatching fuzzing tasks.


\textbf{Seed caching}. 
Since the seeds are relatively heavy-weight and constant, we reduce the seed synchronization overhead via a local cache. We generate a hash value and use it as an index to identify the seed in the database. Then, we maintain a local cache of seeds in the memory of each fuzzing instance. As each fuzzing instance has a seed assigned to it explicitly, when scheduling a task, only the cache miss-hit seeds are retrieved from the database, and the cache hit seeds avoid the seed copy and improve efficiency. Specifically, we modify the functions to read and write seeds in AFL to redirect the seed’s access. First, the local map will be referred to and the seed is only downloaded from the database when it does not exist in the local map. Similarly, the seed writer will upload the seed to the database and maintain a copy on the local map at the same time. We abandon the seed queue for local seeds in each fuzzing instance and move the scheduling scheme to the scheduler. 

\textbf{Coverage broadcast}.
In AFL, coverage information is manifested by a bitmap compressed as high-density raw data (e.g., the 64 KB bitmap in AFL). We do not share coverage information directly among all the fuzzing instances because it changes rapidly and frequently. For example, AFL may alter the bitmap on each execution of a test case. It is easy to cause conflicts when multiple instances update coverage information simultaneously. To alleviate such conflicts, we synchronize coverage information in a centralized manner with two steps. 
We maintain a global bitmap at the main node to receive coverage change from the local instances and broadcast the new coverage information back to them.
Each time a local instance incurs a bitmap change, instead of synchronizing with all the instances, it only notifies the global bitmap. After the global bitmap has been updated, it broadcasts this update to all the local instances. Then, the instances copy the global bitmap to local to finish the update. Notably, when multiple instances update the global bitmap concurrently, the updates are at the byte level. Though there still is a lock/unlock behavior, it avoids locking/unlocking the whole bitmap (64kb) and can significantly lower the chances of bitmap update conflict. Thus, it avoids performance deduction.
Since concurrent reading of the global bitmap does not have conflicts, the broadcast process is fast and efficient.


\textbf{Instant synchronization}.
Unlike the periodical synchronization used in previous work \cite{roving,disafl,PAFL-N,enfuzz,pfuzz1}, our approach achieves instant synchronization. Every time a new seed is discovered and uploaded to the database, along with its fuzzing status, both are instantly accessible to all the fuzzing instances. Every time the global bitmap coverage is updated, the local bitmaps in the fuzzing instances are synchronized accordingly once the undertaking fuzzing tasks are finished.

\subsection{Elastic Seed Evaluation Computing Power Allocation}
\label{dynamic}

In a distributed fuzzing environment, different fuzzing instances can generate duplicate seeds simultaneously, such duplicate seeds include identical seeds and seeds execute the same path. Duplicate seeds lead to redundant execution and waste of testing resources. Thus, we evaluate the new seeds to filter out the duplicates every time new seeds are generated. However, when many new seeds need evaluating, centralized task scheduling can represent a bottleneck that places a heavy burden on seed evaluation. 
Besides, the number of new seeds is not well-distributed. The new seeds can flood into the scheduler within a short interval while remaining at a low volume for the rest of the time. 

To alleviate such limitations, we propose shifting some of the seed evaluation work from the scheduler to professional evaluating instances, and such evaluation instances are elastically adjusted according to the number of seeds to be evaluated. The elasticity of evaluating instances contributes to handling the dramatic increase of seed evaluation requests, which can adaptively coordinate the two groups of instances and achieves a global optimization of resource allocation.

Whenever a fuzzing instance uploads new seeds to the database, the scheduler will receive an ``update'' signal, causing it to initiate an evaluating thread to de-duplicate seeds, which it terminates when there are no new seeds left. However, when too many new seeds flood into the scheduler, the scheduler will alleviate the over-burden by shifting some fuzzing instances to temporarily evaluate these new seeds by dynamically converting them to evaluation instances. The invocation of such an elastically evaluation expanding scheme depends on the number of unevaluated seeds in the scheduler. The scheduler checks the number of unevaluated seeds at intervals and adjusts the number of evaluating instances. A threshold is used to control the interval. Intuitively, a lower threshold will invoke the adjustment easily and frequently, which will waste computing resources. In comparison, a higher threshold will cause new seeds to heap up and depress the overall performance. We empirically set the initial threshold as 1000, which means the scheduler rechecks the number of seeds to be evaluated and adjusts the evaluating instances every time the scheduler has received 1000 ``update'' signals.

\begin{algorithm}
\caption{Elastic seed evaluation computing power allocation.}
\label{alg:expanding}
\algsetup{linenosize=\footnotesize}
\scriptsize
\begin{algorithmic}
\STATE {$updates \leftarrow 0$}
\STATE {$threshold \leftarrow 1000$}
\STATE {$seeds\_to\_evaluate \leftarrow 0$}
\WHILE {True} 
\IF {receive $new\_seed$ from fuzzing instances}
\STATE {$updates \leftarrow updates + 1$} // increase 1 for each new seed.
\STATE {$seeds\_to\_evaluate \leftarrow seeds\_to\_evaluate + 1$} 
\ENDIF
\IF {$updates \geqslant threshold$}
\STATE {$eval\_instance\_num \leftarrow [seeds\_to\_evaluate/evaluate\_speed]$}
\STATE {$unique\_rate \leftarrow total\_evaluated\_seeds/unique\_seeds\_num$}
\IF {$eval\_instance\_num > unique\_rate/2$}
\STATE {$eval\_instance\_num \leftarrow [unique\_rate/2]$}
\ENDIF
\IF {$eval\_instance\_num < 2$}
\STATE {$eval\_instance\_num \leftarrow 2$}
\ENDIF
\FOR {$0 \leqslant $i$ < eval\_instance\_num$}
\STATE{$instance[i].flag \leftarrow ''evaluate''$}
\ENDFOR
\FOR {$eval\_instance\_num \leqslant $i$ < total\_instance\_num$}
\STATE{$instance[i].flag \leftarrow ''fuzz''$}
\ENDFOR
\IF {$seeds\_to\_evaluate > evaluate\_speed * eval\_instance\_num$}
\STATE {$threshold \leftarrow threshold / 2$}
\ELSE
\STATE {$threshold \leftarrow threshold * 2$}
\ENDIF
\STATE {$updates \leftarrow 0$}
\ENDIF
\ENDWHILE
\end{algorithmic}
\end{algorithm}

As Algorithm~\ref{alg:expanding} shows, the number of evaluating instances to be expanded \texttt{eval\_instance\_num}  is estimated by the number of \texttt{seeds\_to\_evaluate} divided by the \texttt{evaluate\_speed}. \texttt{evaluate\_speed} is a dynamic statistical value based on the average number of seeds that have been evaluated in a second. We use \verb:unique_rate: as an estimation of the previous de-duplication performance to expect how many duplicate seeds would be removed, which can help to adjust the number of evaluating instances. Variable \verb:unique_rate: is also used as a heuristic threshold to restrict the number of evaluating instances. Based on our observation, to use resource efficiently, \verb:eval_instance_num: should not be greater than \verb:unique_rate/2:. Otherwise, the evaluating instances would be overplus.
For each evaluating instance that has been expanded, the flag is changed from ``fuzz'' to ``evaluate'', indicating that the instance would request seeds from \verb:seeds_to_evaluate: to evaluate rather \texttt{prioritized\_seed\_queue} to fuzz. Notably, the instance will wait for completing the current fuzzing (or evaluating) task before role switching. 
If the new seeds received in \verb:seeds_to_evaluate: exceeds the \verb:threshold:, the instances would be reallocated. 
Variable \verb:threshold: is used to adaptively control the frequency we re-allocate the instances. Initially, the instances are re-allocated every 1000 updates (i.e.,1000 new seeds received). When there are too many seeds to evaluate, we should shorten the frequency by setting \verb:threshold: to \verb:threshold/2:, so that the evaluating instances would be allocated in time. Otherwise, \verb:threshold: is doubled.

We conduct de-duplication in three steps. First, we use the hash value of seeds to remove identical seeds. Second,  we use the execution path checksum to identify different seeds that exercise the same path. Finally, for seeds that execute the same path but have different checksums, we use the bitmap to compare the coverage. Though this situation is counter-intuitive, our experiments show that the checksum in AFL is environment-dependent. Seeds that execute the same path would have different checksums on different machines. For such seeds, based on the comparison of the bitmaps, we discard new seeds that do not extend coverage and keep the rest. In our approach, the first two steps are simple and suitable for most cases, while the third step is expensive but can refine de-duplication at a deeper level.

\section{Implementation}
Based on the design of centralized dynamic scheduling, we implement a distributed fuzzing tool called UltraFuzz. UltraFuzz was built on top of AFL (version 2.52b) by adding 3500 lines of C code. The scheduler, which also controls instances to alter between fuzzing and evaluating, is implemented in the C language. It communicates with instances based on TCP, which is realized by the socket and directs instances to perform either fuzzing or evaluating tasks. Since the socket communication based on TCP is blocking, we use {\emph{select} (an interface of a system call) to handle the concurrent requests from multi-instances, which performs better in current processing with non-blocking. 
The database is built on MongoDB, which is used to store seeds and their fuzzing status. All of the other parts, including the scheduler, fuzzing instances, and evaluating instances, use \emph{libmongoc} to interact with the database. The fuzzing instances are implemented based on AFL. The evaluating instance is self-implemented via C language.
UltraFuzz has been made publicly available online \footnote{\url{https://gitlink.org.cn/hunter-2018/Ultrafuzz.git}} to encourage and support future research. 


\section{Evaluation}
In this section, we evaluate UltraFuzz with real-world experiments, aiming to answer the following research questions:
(1) Whether the techniques adopted in UltraFuzz are effective in resource-saving?
(2) What can we benefit from resource-saving in distributed fuzzing?

\textbf{Target Programs.}
We used 16 widely adopted real-world programs to run the tests, as shown in Table \ref{table:config}. 
To facilitate the comparison of different tools in large-scale parallel fuzzing before coverage saturation occurs,
we only select programs that are relatively large. 
For programs from the Google Fuzzer-test-suite, we use the initial seeds from the test suite as the initial seeds. For the rest programs, we use seeds in the \texttt{testcase} directory provided by AFL as the initial seeds. 

\begin{table}[hbtp]


\setlength{\abovecaptionskip}{0cm}
\scriptsize
\centering
\caption{The configuration of programs under test}
\label{table:config}
\begin{tabular}{ccccc}
\toprule
\textbf{Subjects}&\textbf{Version}&\textbf{Format}&\textbf{Size}&\textbf{LoC}\\
\midrule
boringssl @@& 2016-02-12 & lib&6.8M  &0.3k\\
freetype @@&2017&font&6.3M           &0.5k \\
libcxx @@&2017-01-27 & lib&1.9M       &5.0k\\
libxml @@&libxml2-v2.9.2  & xml&12M     &15.7k\\
re2 @@&2014-12-09 & lib&5.6M              &0.9k\\
libarch @@& libarch 2017-01-04& text&3.7M     &3.0k\\
size @@ & Binutils-2.34 &elf&10M         &7.9k\\
readelf -a @@ & Binutils-2.34 &elf&5.4M          &20.5k\\
objdump -d @@ &Binutils-2.34 &elf&16M       &5.4k\\
avconv -y -i @@ -f null & Libav-12.3 & mp4&77M     &2.9k\\
infotocap @@ & ncurses-6.1 &text &1.1M          &4.9k\\
pdftotext @@ /dev/null & xpdf-4.02 &pdf&7.9M         &0.9k\\
tiff2bw @@ /dev/null & tiff-4.1 &tiff &2.6M          &0.5k\\
ffmpeg -i @@&ffmpeg-4.1.3 &mp4 &41M        &4.9k\\
gnuplot @@ &gnuplot-5.5 &text&8.5M          &1.0k\\
tcpdump -nr @@&tcpdump-4.9.3 &pcap&6.3M          &2.6k\\
\bottomrule
\end{tabular}
\end{table}

\textbf{Compared tools.}
To thoroughly evaluate the performance of UltraFuzz, we choose to compare UltraFuzz with AFL, PAFL, EnFuzz, and AFL-P. We select AFL as the baseline. For tools that do not support the inter-machine mode, such as AFL-P,  PAFL, and EnFuzz, we virtually expand the experiment scale by extending the test time. We propose a criterion called \textit{computing resource}, which is the product of computing cores and testing hours. We define one unit of computing resource as a core multiplied by an hour. In this way, for tools that do not support inter-machine mode, such as AFL, when the requested testing cores are beyond a single machine, we extend the fuzzing time to compensate for the cores. For example, a 128-core 1-hour test can be equally replaced by a 16-core machine running for 8 hours. To fairly evaluate the influence caused by inter-machine communication, we also include the machine number when physically expanding the experiment scale. Namely, the machine number should be the experiment scale divided by 16 (for a 32-core machine, we only use half the cores to avoid potential resource interference). For example, for testing that requires 128 cores, we use 8 machines with each machine contributing 16 cores. Specifically, the tests of UltraFuzz are physically expanded on machines with 16 cores occupied, and the tests of the rest tools are virtually expanded.


\begin{table}
\setlength{\abovecaptionskip}{0cm}
  \centering  \scriptsize
  \caption{Computing resource distribution}
  \label{table:distribution}
    \begin{tabular}{l|rrrrr}
   \toprule
 Tools   &8 units  &16 units &32 units &64 units &128 units  \\
   \midrule
AFL &1m1c8h &1m1c16h &1m1c32h &1m1c64h &1m1c128h\\
AFL-P  &1m8c1h &1m16c1h &1m16c2h &1m16c4h &1m16c8h\\
PAFL &1m8c1h &1m16c1h &1m16c2h &1m16c4h &1m16c8h\\
EnFuzz  &1m8c1h &1m16c1h &1m16c2h &1m16c4h &1m16c8h\\
UltraFuzz  &1m8c1h &1m16c1h &2m16c1h &4m16c1h &8m16c1h\\
   \bottomrule
    \end{tabular}  
    \begin{tablenotes}  
\item[*] e.g., 4m16c1h means using 4 machines with each 16 cores running for 8 hours.
\end{tablenotes} 
\end{table}

\textbf{Configuration.}
We use UltraFuzz, PAFL, EnFuzz, and AFL-P to fuzz each program for 8, 16, 32, 64, and 128 units of computing resources, respectively. Each experiment is repeated \textbf{10} times to reduce randomness. Particularly, we select the single mode of AFL to run for  8, 16, 32, 64, and 128 hours as the baseline. 
The campaigns distribution are listed in Table \ref{table:distribution}.
\textbf{To clarify, though UltraFuzz runs tests only for one hour, it uses many cores running concurrently to share the task and shorten the time. The real test time should multiply the number of cores to measure the capacity of work. Besides, the baseline tests can validate the tests.}
Our experiments were conducted on eight machines of Intel(R) Xeon(R) CPU E5-2620 v4 @ 2.10GHz with 32 cores, running a 64-bit CentOS Linux release 7.7.1908 (Core). 
It is worth mentioning that, \textbf{to avoid the potential computing resource interference within a machine \cite{bohmefuzzing}, for each machine, we only use half of the cores (i.e., 16) for fuzzing and leave the rest unoccupied. Thus, the experiment scale is also the time of the machines}. Moreover, when the experiment scale is beyond 16, we can guarantee that the testing environment is inter-machine and equally distributed. No experiment is running concurrently. We eliminate the unnecessary processes in the background of each machine to keep the machine usage stay as equal as possible.
Notably, the total time of our experiments is more than 6,186 CPU days.

\subsection{Effectiveness of Resource-saving}
Given the same units of computing resources, intuitively, a resource-saving fuzzer consumes less resources in finding a new path and thus has larger coverage. Here we use two criteria to evaluate the effectiveness of resource-saving. Except for the classical coverage metric, we also define \textit{path cost} as a metric to measure the resource efficiency of the fuzzer.

\subsubsection{Coverage}
\label{branch}

\begin{figure*}[h]
\setlength{\abovecaptionskip}{0cm}
  \centering
  \includegraphics[width=\linewidth]{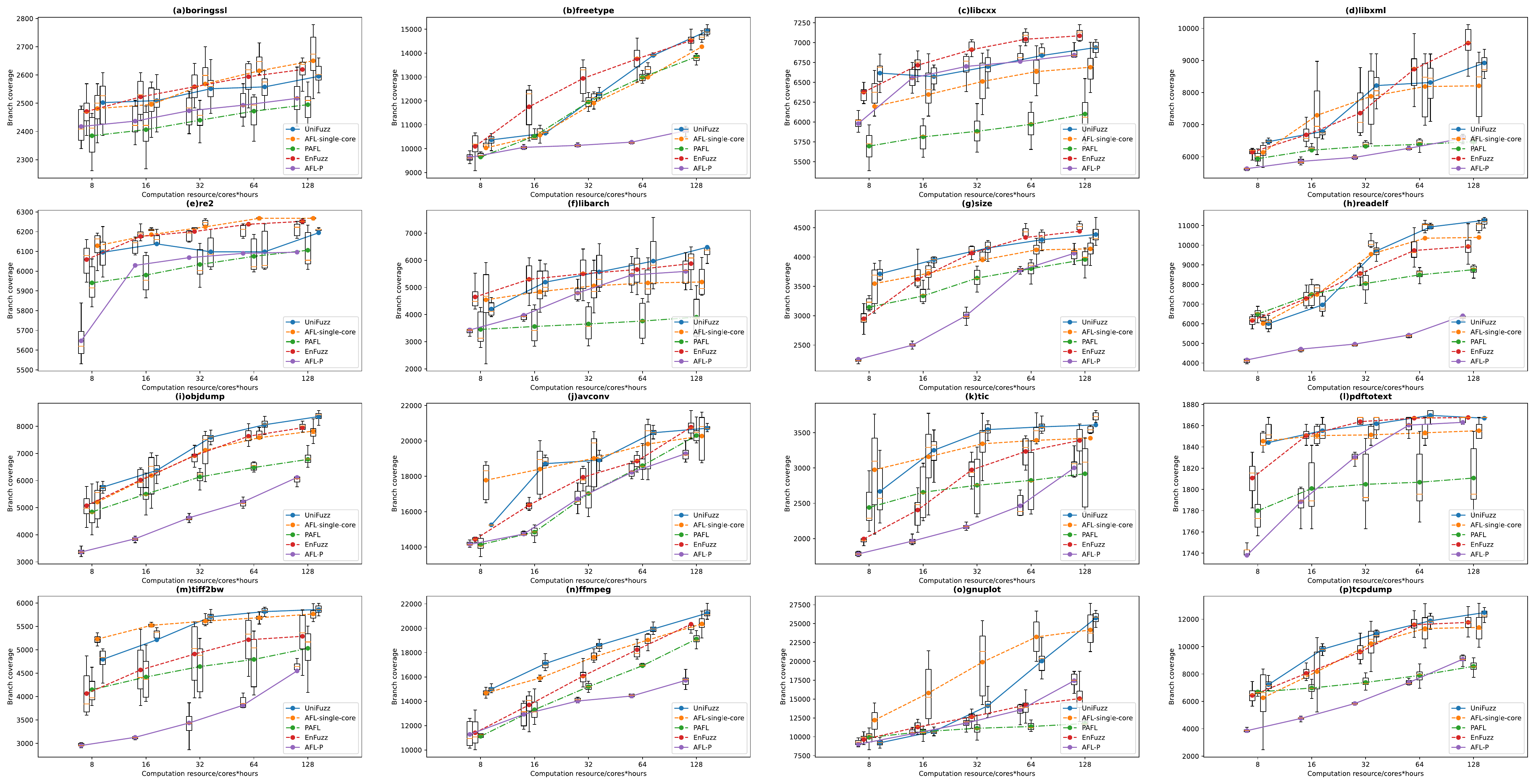}
  \caption{Comparison of branch coverage reached by different tools with the same computing resources.}
  \label{fig:branchcoverage}
\end{figure*}

\begin{table*}
  \centering \scriptsize
  \caption{Branch coverage / Test cases increment of UltraFuzz compared to baseline with same computing resources.}
  \label{table:resource}
    \begin{tabular}{l|rrrrrr}
   \toprule
   \multirow{2}{*}{{Program}} & \multicolumn{1}{c}{{8 units}}  & \multicolumn{1}{c}{{16 units}}   & \multicolumn{1}{c}{{32 units}}   & \multicolumn{1}{c}{{64 units}}   & \multicolumn{1}{c}{{128 units}}         \\ \cline{2-6} 
& {Branch coverage/Test cases} & {Branch coverage/Test cases} & {Branch coverage/Test cases} & {Branch coverage/Test cases} & Branch coverage/Test cases  \\
   \midrule
boringssl&2,502(1.01x)/30.8M(1.23x)&2,509(1.01x)/50.7M(1.09x)&2,551(0.99x)/44.3M(0.44x)&2,557(0.98x)/75.6M(0.38x)&2,595(0.98x)/121.0M(0.33x)\\
freetype&10,368(1.03x)/23.1M(1.06x)&10,646(1.01x)/36.4M(0.86x)&12,293(1.03x)/100.4M(1.22x)&13,905(1.07x)/190.6M(1.17x)&14,923(1.05x)/345.4M(1.08x)\\
libcxx&6,616(1.07x)/45.6M(1.56x)&6,572(1.04x)/42.0M(0.95x)&6,698(1.03x)/53.4M(0.79x)&6,837(1.03x)/73.1M(0.59x)&6,938(1.04x)/126.9M(0.60x)\\
libxml&6,469(1.05x)/30.3M(1.01x)&6,788(0.93x)/44.9M(0.78x)&8,216(1.04x)/103.4M(1.18x)&8,313(1.02x)/165.9M(1.34x)&8,921(1.09x)/256.8M(1.85x)\\
re2&6,095(0.99x)/19.6M(0.55x)&6,138(0.99x)/24.8M(0.41x)&6,098(0.98x)/37.3M(0.28x)&6,098(0.97x)/50.8M(0.17x)&6,194(0.99x)/73.1M(0.22x)\\
libarch&4,204(0.93x)/19.1M(1.01x)&5,193(1.07x)/44.8M(1.42x)&5,570(1.10x)/78.5M(1.45x)&5,970(1.16x)/157.4M(1.64x)&6,481(1.25x)/278.8M(2.62x)\\
size&3,713(1.05x)/27.2M(1.03x)&3,951(1.06x)/45.1M(0.95x)&4,152(1.05x)/97.9M(1.11x)&4,293(1.04x)/161.4M(0.98x)&4,382(1.06x)/242.1M(1.31x)\\
readelf&6,004(1.00x)/31.5M(0.78x)&6,961(0.93x)/58.1M(0.77x)&9,707(1.02x)/143.7M(1.01x)&10,931(1.06x)/246.5M(0.95x)&11,265(1.08x)/357.8M(1.35x)\\
objdump&5,759(1.11x)/24.7M(1.10x)&6,373(1.03x)/40.1M(0.96x)&7,594(1.07x)/95.6M(1.19x)&8,060(1.06x)/170.7M(1.20x)&8,360(1.07x)/242.9M(1.10x)\\
avconv&15,255(0.86x)/2.0M(0.90x)&18,714(1.02x)/6.7M(1.63x)&18,912(0.99x)/10.3M(1.28x)&20,461(1.03x)/22.6M(1.42x)&20,714(1.02x)/33.5M(1.64x)\\
infotocap&2,667(0.90x)/12.1M(1.21x)&3,251(1.03x)/21.1M(1.26x)&3,544(1.06x)/38.4M(1.52x)&3,582(1.06x)/67.2M(1.61x)&3,609(1.05x)/97.6M(2.08x)\\
pdftotext&1,844(1.00x)/6.1M(0.81x)&1,855(1.00x)/11.1M(0.91x)&1,861(1.01x)/22.8M(1.30x)&1,869(1.01x)/44.1M(1.66x)&1,867(1.01x)/82.1M(1.86x)\\
tiff2bw&4,796(0.92x)/38.1M(0.78x)&5,213(0.94x)/60.8M(0.69x)&5,704(1.02x)/115.8M(0.78x)&5,816(1.02x)/181.2M(0.84x)&5,860(1.02x)/252.4M(0.92x)\\
ffmpeg&15,000(1.02x)/6.7M(0.92x)&17,096(1.07x)/14.8M(1.09x)&18,604(1.05x)/29.5M(1.12x)&19,935(1.05x)/61.4M(1.20x)&21,240(1.04x)/112.1M(1.39x)\\
gnuplot&9,205(0.75x)/13.1M(0.77x)&10,716(0.68x)/27.2M(0.88x)&14,110(0.71x)/62.2M(1.15x)&20,043(0.86x)/110.1M(1.32x)&25,740(1.07x)/184.4M(1.91x)\\
tcpdump&7,219(1.15x)/36.6M(1.39x)&9,805(1.20x)/76.3M(1.53x)&10,942(1.07x)/115.0M(1.25x)&11,882(1.05x)/175.2M(1.18x)&12,483(1.1x)/294.8M(1.86x)\\
\midrule
Average &0.99x / 1.01x&1.00x / 1.01x&1.01x / 1.07x&1.03x / 1.1x&1.06x / 1.38x\\
   \bottomrule
    \end{tabular}  
    \begin{tablenotes}  
\item[*] B/T means the branch coverage / number of test cases generated by UltraFuzz. And the value in bracket represents the corresponding increment of UltraFuzz compared to AFL.
\end{tablenotes} 
\end{table*}

\begin{table*}
\setlength{\abovecaptionskip}{0cm}
  \centering  \scriptsize
  \caption{The standard deviations and 95\% confidence intervals of branch coverage reached by UltraFuzz}
  \label{table:confidence intervals}
    \begin{tabular}{l|rrrrrr}
   \toprule
   \multirow{2}{*}{{Program}} & \multicolumn{1}{c}{{8 units}}  & \multicolumn{1}{c}{{16 units}}   & \multicolumn{1}{c}{{32 units}}   & \multicolumn{1}{c}{{64 units}}   & \multicolumn{1}{c}{{128 units}}         \\ \cline{2-6} 
& {Deviation/Interval} &{Deviation/Interval}& {Deviation/Interval} &{Deviation/Interval}& {Deviation/Interval}  \\
   \midrule

boringssl&78.63 / 2,348-2,656 &56.1 / 2,399-2,619 &53.89 / 2,445-2,656&47.71 / 2,464-2,651&46.6 / 2,503-2,686\\
freetype&205.88 / 9,965-10,772&138.47 / 10,375-10,918&160.36 / 11,978-12,607&101.98 / 13,705-14,105&160.17 / 14,609-15,237\\
libcxx&148.67 / 6,325-6,907&113.03 / 6,350-6,794&154.37 / 6,395-7,000&150.54 / 6,542-7,132&71.66 / 6,798-7,079\\
libxml&82.78 / 6,307-6,631&215.07 / 6,366-7,209&532.65 / 7,172-9,260&686.27 / 6,968-9,658&426.43 / 8,085-9,757\\
re2&79.05 / 5,940-6,250&66.33 / 6,008-6,268&78.82 / 5,943-6,252&97.13 / 5,907-6,288&47.15 / 6,101-6,286\\
libarch&251.7 / 3,710-4,697&427.78 / 4,355-6,032&659.24 / 4,278-6,862&870.31 / 4,265-7,676&561.63 / 5,380-7,582\\
size&110.83 / 3,496-3,931&142.45 / 3,671-4,230&113.74 / 3,930-4,375&166.38 / 3,967-4,619&144.11 / 4,100-4,665\\
readelf&270.34 / 5,475-6,534&445.27 / 6,088-7,833&257.63 / 9,202-10,212&557.6 / 9,838-12,024&322.19 / 10,633-11,896\\
objdump&171.44 / 5,423-6,095&296.93 / 5,791-6,955&172.44 / 7,256-7,932&192.71 / 7,682-8,437&169.32 / 8,029-8,692\\
avconv&35.15 / 15,186-15,324&452.95 / 17,826-19,601&436.43 / 18,057-19,768&218.28 / 20,033-20,889&160.7 / 20,399-21,029\\
infotocap&341.25 / 1,999-3,336&207.21 / 2,845-3,657&83.3 / 3,380-3,707&73.51 / 3,438-3,726&378.88 / 2,867-4,352\\
pdftotext&34.33 / 1,776-1,911&6.84 / 1,841-1,868&8.0 / 1,846-1,877&4.2 / 1,861-1,877&3.53 / 1,860-1,874\\
tiff2bw&213.62 / 4,377-5,215&356.58 / 4,514-5,912&83.27 / 5,541-5,868&86.66 / 5,647-5,986&89.43 / 5,685-6,036\\
ffmpeg&219.62 / 14,569-15,430&461.28 / 16,192-18,000&398.98 / 17,822-19,386&272.01 / 19,402-20,468&400.29 / 20,455-22,025\\
gnuplot&393.9 / 8,433-9,977&422.27 / 9,889-11,544&1235.13 / 11,690-16,531&1567.88 / 16,970-23,116&760.88 / 24,248-27,231\\
tcpdump&339.79 / 6,554-7,885&245.8 / 9,323-10,287&492.32 / 9,977-11,907&378.15 / 11,141-12,623&665.16 / 11,179-13,786\\

   \bottomrule
    \end{tabular}  
\end{table*}

By saving computing resources, a fuzzer can run more tests and potentially enlarge code coverage. 
To avoid the potential influence caused by order-dependent of path coverage in AFL implementation \cite{fairfuzz}, 
we use branch coverage as the main criterion to measure the performance of these fuzzers. However, we also provide the path coverage results in the Appendix (Fig. \ref{fig:pathcoverage}) for reference, which is consistent with branch coverage results.

Fig.~\ref{fig:branchcoverage} plots the average branch coverage discovered by these tools throughout 10 runs with different computing resources. We can see that the branch coverage reached by each tool rises as computing resources increase on most programs. Among these tools, UltraFuzz reaches the highest branch coverage in most (10/16) programs, 
outperforming the other four tools. In particular, when compared with the baseline AFL-single,  UltraFuzz performs better on 14 programs with the same computing resources, such as \emph{freetype}, \emph{libcxx}, \emph{size} and \emph{readelf}. However, the other four tools do not reach a higher branch coverage than that of AFL-single. AFL-single discovers more branches than PAFL and AFL-P on almost all programs using the same computing resources. Compared with EnFuzz, AFL-single achieves similar performance on these programs.

Notable is, PAFL and EnFuzz have the advantage over UltraFuzz in two aspects. First, PAFL and EnFuzz optimize the scheduling algorithm or mutation strategy, which makes them more efficient than UltraFuzz in schedule and mutation. Second, though based on an equal computing resource comparison, PAFL and EnFuzz run with fewer cores but more time than UltraFuzz due to the intra-machine limitation, which alleviates the overhead of large-scale parallelism. From this result, we infer the reason that UltraFuzz outperforms the other tools lies in the optimization of the parallel architecture.

\begin{table}[h]
\setlength{\abovecaptionskip}{0cm}
  \centering  \scriptsize
  \caption{The p-value results of UltraFuzz regarding branch coverage.}
  \label{table:statistical2}
    \begin{tabular}{l|cccc}
   \toprule
      Program &$p_1$(\&AFL)  &$p_2$(\&PAFL) &$p_3$(\&EnFuzz) &$p_4$(\&AFL-P) \\
   \midrule
boringssl&$2.1*10^{-1}$&$3.0*10^{-4}$&$2.5*10^{-1}$&$6.6*10^{-3}$\\
freetype&$1.6*10^{-1}$&$2.3*10^{-9}$&$2.0*10^{-3}$&$4.9*10^{-22}$\\
libcxx&$3.5*10^{-3}$&$3.5*10^{-9}$&$5.0*10^{-3}$&$5.6*10^{-2}$\\
libxml&$3.9*10^{-2}$&$6.4*10^{-12}$&$9.1*10^{-3}$&$3.8*10^{-12}$\\
re2&$2.2*10^{-4}$&$1.4*10^{-2}$&$2.0*10^{-3}$&$2.8*10^{-1}$\\
libarch&$8.9*10^{-5}$&$1.7*10^{-7}$&$2.8*10^{-2}$&$1.9*10^{-3}$\\
size&$1.9*10^{-3}$&$1.1*10^{-5}$&$6.0*10^{-1}$&$5.3*10^{-5}$\\
readelf&$1.9*10^{-1}$&$4.3*10^{-11}$&$1.4*10^{-5}$&$1.8*10^{-18}$\\
objdump&$5.7*10^{-4}$&$2.6*10^{-10}$&$4.3*10^{-4}$&$2.6*10^{-14}$\\
avconv&$2.9*10^{-1}$&$6.4*10^{-2}$&$7.1*10^{-1}$&$4.3*10^{-9}$\\
infotocap&$2.8*10^{-1}$&$1.6*10^{-3}$&$1.4*10^{-1}$&$5.9*10^{-4}$\\
pdftotext&$2.3*10^{-4}$&$2.2*10^{-5}$&$5.9*10^{-1}$&$1.9*10^{-1}$\\
tiff2bw&$5.6*10^{-2}$&$2.9*10^{-5}$&$2.7*10^{-3}$&$1.2*10^{-7}$\\
ffmpeg&$1.9*10^{-3}$&$1.9*10^{-9}$&$2.9*10^{-2}$&$3.7*10^{-16}$\\
gnuplot&$5.1*10^{-2}$&$4.4*10^{-18}$&$2.2*10^{-12}$&$1.5*10^{-10}$\\
tcpdump&$1.6*10^{-2}$&$5.1*10^{-11}$&$2.7*10^{-2}$&$2.6*10^{-11}$\\
   \bottomrule
    \end{tabular} 
\end{table}

Furthermore, following the guidance of \cite{klees2018evaluating}, we conducted statistical analysis to measure the gap in performance between different tools by calculating p-values about branch coverage. Considering the randomness of fuzzing, we conduct the two independent-sample t-test to analyze the results. 
Specifically, by Kolmogorov-Smirnov (K-S) test, we firstly confirmed that almost every group of the numbers of branches in repeated testing is normally distributed. Then we utilized the Levene test to confirm that the variances of different groups are very close. Based on these steps, we assume that each group of data is normally distributed with the approximate variance, then we conducted the two independent-sample t-test to calculate the p-value.

As Table \ref{table:statistical2} lists, $p_1$, $p_2$, $p_3$, and $p_4$ represent those differences between the performances of UltraFuzz and AFL, PAFL, EnFuzz, and AFL-P, respectively.  From the results, for branch coverage, we can conclude that UltraFuzz outperforms PAFL and AFL-P significantly as $p_3$ and $p_4$ are smaller than $10^{-4}$ on most programs. Compared to EnFuzz and AFL, UltraFuzz performs better on many programs (e.g., \emph{libarch}, \emph{objdump}, \emph{ffmpeg} and \emph{tcpdump}) with that $p_1$ and $p_2$ are smaller than $10^{-2}$. Moreover, we also calculate the standard deviations and 95\% confidence intervals of branch coverage reached by UltraFuzz during 10 repeated tests, which are listed in Table \ref{table:confidence intervals}. From Table \ref{table:confidence intervals}, the standard deviations of coverage reached by UltraFuzz with 128 units of computing resource on most (11/16) programs are less than 400. Particularly, on the programs where the average branches covered by UltraFuzz are no more than 10,000 (e.g., \emph{boringssl}, \emph{libcxx}, \emph{objdump}, \emph{pdftotext}), UltraFuzz performs steadily with the small standard deviations and short confidence intervals.
In conclusion, UltraFuzz outperforms the other four tools in branch coverage and performs steadily with 128 units of computing resource on most programs.

\subsubsection{Path Cost}
\label{path_cost}

\begin{figure}[h]
\setlength{\abovecaptionskip}{0cm}
  \centering
  \includegraphics[width=0.8\linewidth]{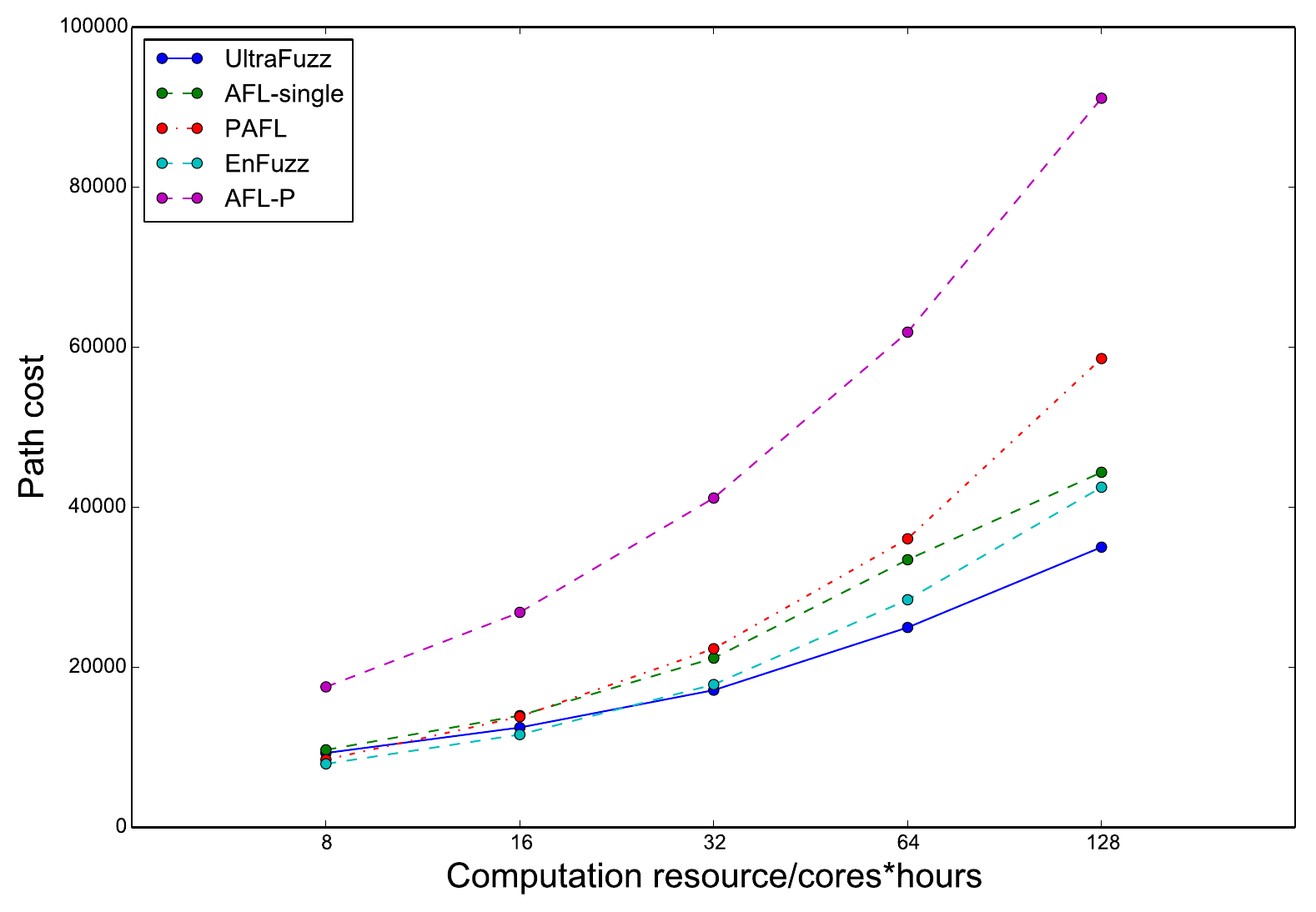}
  \caption{Comparison of different tools on average path cost.}
  \label{fig:average_energy}
\end{figure}

We define \textit{path cost} as the number of generated test cases that divides the number of discovered paths in the campaign, which means the average number of test cases consumed on finding each path. Path cost can roughly reflect the quality of the test cases and the resource efficiency of the fuzzer.

Path cost is calculated on all 16 test programs. To save space, we use Fig. \ref{fig:average_energy} to show their average values. The path cost of each tool rises as computing resources increase. AFL-P has the highest path cost because the poor synchronization scheme in AFL-P makes it produce large numbers of redundant test cases and thus has low resource efficiency. UltraFuzz has the lowest path cost, which is lower than the baseline AFL-single. We also provide the result of path cost regarding each test program in the Appendix (Fig. \ref{fig:total_average_energy}), which is consistent with the above description.
Among these tools, AFL-P has the highest path cost in most (13/16) programs. UltraFuzz achieves the lowest path cost in 6 out of 16 programs, which is similar to the baseline AFL-single but better than the other tools.

Therefore, we can conclude that, by adopting the resource-saving techniques, UltraFuzz has better resource efficiency and averagely consumes fewer test cases to find a new path in contrast to other tools. Consequently, UltraFuzz outperforms the other tools on both path coverage and branch coverage.

\subsection{Super-linear Performance Acceleration}
\label{super}

As for the performance of acceleration, we conduct a close comparison between UltraFuzz and AFL-single on branch coverage. We calculate the improved ratio of the branch coverage reached by UltraFuzz and that reached by AFL-single. As the left side of each column in Table \ref{table:resource} shows, UltraFuzz achieves higher branch coverage than AFL-single on most programs from 8 units to 128 units. Moreover, as computing resources increase, the average speedup of UltraFuzz to AFL-single increases from 0.99x to 1.06x. It is noteworthy that when measured with path coverage, the superiority persists, which is even better, ranging from 1.03x to 1.19x (Table \ref{table:path coverage}).  Namely, we had an empirical observation: In the experiments, UltraFuzz on n cores for 1 hour performs better than AFL on 1 core for n hours, and this improvement increases with the number of cores. For short, we call this phenomenon ``super-linear acceleration''.

We give a fine-grained classification of the parallel fuzzing acceleration as follows.

\begin{itemize}

\item \textbf{Linear acceleration}: The result of AFL-single using 1 core to run for n hours equals a fuzzer using n cores running for 1 hour. This is an ideal situation of parallel fuzzing which excludes the overhead of parallelism and the affection of fuzzing status synchronization.
\item \textbf{Sub-linear acceleration}: The result of AFL-single using one core to run for n hours is better than a fuzzer using n cores running for 1 hour. This is a typical situation, and a known example is the parallel mode of AFL (i.e., AFL-P).
\item \textbf{Super-linear acceleration}: The result of AFL-single using one core to run for n hours is worse than a fuzzer using n cores running for 1 hour. An example is UltraFuzz.

\end{itemize}

Generally speaking, if we only expand AFL to a parallel mode without optimizing its scheduling, theoretically, the parallel mode will not perform as well as AFL-single with the same computing resources. The reason is that parallelism always introduces additional overhead, such as by seeds synchronization and task dispatching. 
Hence, with the same computing resources, the proportion used on fuzzing seeds in parallel mode is less than those in AFL-single. As a result, the test cases produced by AFL’s parallel mode should be less than those produced by extending the execution time of AFL-single. Based on this assumption, we analyzed the average number of test cases generated by UltraFuzz and AFL-single, as listed on the right side of each column in Table \ref{table:resource}. However, the results show that, on average, UltraFuzz generates the same or more test cases than AFL  single with the same computing resources, which is not consistent with our original inference. We argue the main reason behind this result is the high resource efficiency of UltraFuzz. UltraFuzz consumes fewer resources in generating each valid test case. To dig into the deep reason, we propose the following explanations.

\subsubsection{Branch Coverage for 24 Hours}

\begin{table*}
  \centering \scriptsize
  \caption{Branch coverage increment of UltraFuzz compared to AFL running for 24 hours.}
  \label{table:branch24}
    \begin{tabular}{l|rrrr}
   \toprule
   \multirow{2}{*}{{Program}} & \multicolumn{1}{c}{{8 cores}}  & \multicolumn{1}{c}{{16 cores}}   & \multicolumn{1}{c}{{32 cores}}   & \multicolumn{1}{c}{{64 cores}}    \\ 
   \cline{2-5} 
            & {AFL / UltraFuzz} & {AFL / UltraFuzz} & {AFL / UltraFuzz} & {AFL / UltraFuzz}  \\
   \midrule
boringssl  & 2,828  / 2,657 (0.94x) &2,747 /2,958 (1.08x) &3,132 / 2,661 (0.85x) &3,078 /2,872 (0.93x) \\
freetype   & 11,722 / 13,762 (1.17x) &12,010 / 16,312 (1.36x) &12,146 / 16,485 (1.36x) &12,870 / 16,638 (1.29x) \\
libcxx     & 7,085  / 7,194 (1.02x) &7,220 / 7,367 (1.02x) &7,269 / 7,400 (1.02x) &7,341 /7,407 (1.01x) \\
libxml     & 9,576 /8,361 (0.87x) &9,964 /8,633 (0.87x) &10,042/8,783 (0.87x) &10,598 / 9,832 (0.93x) \\
re2        & 6,313 / 6,218 (0.98x) &6,344 / 6,260 (0.99x) &6,352 / 6,272 (0.99x) &6,371 / 6,286 (0.99x) \\
libarch    & 6,575 / 5,505 (0.84x) &6,794 / 6,193 (0.91x) &7,616 / 7,002 (0.92x) &7,700 / 7,942 (1.03x) \\
size       & 4,336 / 4,724 (1.09x) &4,407 / 4,881 (1.11x) &4,476 / 4,979 (1.11x) &4,692 / 5,034 (1.07x)  \\
readelf    & 9,584 / 11,338 (1.18x) &9,908 / 12,089(1.22x) &9,808 / 12,276 (1.25x) &10,523 / 12,385 (1.18x)  \\
objdump    & 7,736 / 8,471 (1.10x) &7,902 / 8,661 (1.10x) &7,935 / 8,788 (1.11x) &8,123 / 8,878 (1.10x) \\
avconv     & 20,298 / 22,328 (1.10x) &21,092 / 25,384 (1.20x) &21,738 / 26,624 (1.22x) &22,811 / 31,034 (1.36x) \\
infotocap  & 3,353 / 3,654 (1.09x) &3,641 / 3,725 (1.02x) &3,727 / 3,750 (1.01x) &3,714 / 3,766 (1.01x)  \\
pdftotext  & 1,857 / 1,866 (1.00x) &1,862 / 1,870  (1.00x) &1,867  /1,874 (1.00x) &1,871 / 2,013 (1.08x)  \\
tiff2bw    & 5,786 / 6,170 (1.07x) &5,988 / 6,361 (1.06x) &6,078 / 6,557 (1.08x) &6,058 / 6,792 (1.12x)  \\
ffmpeg     & 18,080 / 22,650 (1.25x) &19,049 / 25,613 (1.34x) &19,358 / 27,597 (1.43x) &19,269 / 34,214 (1.78x) \\
gnuplot    & 23,817 / 18,829 (0.79x) &26,747 / 24,602 (0.92x) &27,205 / 30,855 (1.13x) &28,461 / 33,498 (1.18x) \\
tcpdump    & 8,015 /13,529 (1.69x) &10,079 / 14,259 (1.41x) &12,271 / 15,569 (1.27x) &14,865 / 16,936 (1.14x)  \\
\midrule
Average increase  &1.07x        &1.10x              &1.10x             &1.14x\\
   \bottomrule
    \end{tabular}  
    \begin{tablenotes}  
\item[*] The value in bracket represents the corresponding increment of UltraFuzz compared to AFL.
\end{tablenotes} 
\end{table*}

To further analyze the super-linear acceleration, we compared UltraFuzz with AFL for physical scaling across machines (8, 16, 32, and 64 cores, respectively). We run each campaign 10 times for 24 hours to account for the potential difference at the beginning of the campaign. AFL instances run for the same time (24h) across machines but are unsynchronized, where coverage is measured across all corpora.
As Table \ref{table:branch24} lists, compared to AFL, UltraFuzz's branch coverage increments range from 0.79x to 1.78x., and UltraFuzz outperforms AFL for most of the cases (45/64). Averagely, the increments for 8, 16, 32, and 64 cores are 1.07x, 1.10x, 1.10x, and 1.14x, respectively. Moreover, the results also imply that the acceleration gets obvious as the number of cores increases. More specifically, for the group of 8 cores, UltraFuzz outperforms AFL on 10 out of the 16 cases with an average increment of 1.07x. However, for the 64-core group, UltraFuzz outperforms AFL in 13 cases with an average increment of 1.14x. The other two groups (i.e, 16 and 32) are in the middle.
In summary, UltraFuzz still outperforms AFL and the super-linear acceleration phenomenon persists even when the testing time is extended to 24 hours.

\subsubsection{Number of Completed FuzzingTasks }

\begin{table*}
  \centering \scriptsize
  \caption{Number of tasks completed by UltraFuzz compared to AFL running for 24 hours.}
  \label{table:task}
    \begin{tabular}{l|rrrr}
   \toprule
   \multirow{2}{*}{{Program}} & \multicolumn{1}{c}{{8 cores}}  & \multicolumn{1}{c}{{16 cores}}   & \multicolumn{1}{c}{{32 cores}}   & \multicolumn{1}{c}{{64 cores}}    \\ 
   \cline{2-5} 
            & {AFL / UltraFuzz} & {AFL / UltraFuzz} & {AFL / UltraFuzz} & {AFL / UltraFuzz} \\
   \midrule
boringssl &175,090 / 128,416 (0.73x)  &247,020 / 417,671 (1.69x)   &551,606 / 1,426,948 (2.59x)  &643,018 / 2,102,387 (3.27x) \\
freetype  &2,523 / 1,476  (0.59x)     &5,082 / 25,585 (5.03x)      &8,358 / 60,510 (7.24x)     &15,842 / 119,622 (7.55x)    \\
libcxx    &62,282 / 77,781 (1.25x)   &141,466 / 277,750 (1.96x)   &307,524 / 444,015 (1.44x)  &474,077 / 863,904 (1.82x)   \\
libxml    &9,578 / 5,897 (0.62x)      &21,290 / 16,929 (0.80x)     &34,806 / 30,056 (0.86x)    & 63,689 / 61,297 (0.96x)     \\
re2       &179,255 / 364,646 (2.03x) &393,940 / 1,377,637 (3.50x) &520,728/2,218,812 (4.26x) &1,163,060 / 5,198,349 (4.47x) \\
libarch   &8,275 / 3,399 (0.41x)     &16,016 / 15,193 (0.95x)     &32,112 / 36,711 (1.14x)   &64,384 / 89,128 (1.38x)     \\
size      &27,408 / 45,894 (1.67x)   &45,291 / 122,013 (2.69x)    &67,599 / 519,371 (7.68x)  &109,878 / 1,054,117 (9.59x) \\
readelf   &13,271 / 7,639 (0.58x)    &25,455 / 37,881 (1.49x)     &42,722 / 79,609 (1.86x)   &88,223 / 137,259 (1.56x)    \\
objdump   &9,123 / 14,387 (1.58x)    &18,670 / 402,39 (2.16x)     &32,557 / 56,377 (1.73x)   &58,498 / 78,909 (1.35x)     \\
avconv    &136 / 141 (1.04x)        &252 / 622 (2.47x)          &536 / 1,400 (2.61x)      &1,151 / 2,167 (1.88x)       \\
infotocap &2,951 / 1,084 (0.37x)    &6,523 / 1,500 (0.23x)      &12,199 / 2,842 (0.23x)   &22,079 / 7,000 (0.32x)      \\
pdftotext &2,504 / 5,243 (2.09x)     &4,703 / 16,239 (3.45x)     &7,546 / 28,625 (3.79x)   &16,830 / 153,019 (9.09x)    \\
tiff2bw   &12,31 / 57,484 (4.67x)   &25,126 / 114,491 (4.56x)   &41,772 / 476,314 (11.40x) &80,309 / 505,215 (6.29x)   \\
ffmpeg    &648 / 646 (1.00x)        &1,417 / 1,588 (1.12x)      &2,578 / 2,340 (0.91x)    &3,826 / 5,093 (1.33x)      \\
gnuplot   &29,095 / 3,275 (0.11x)   &58,072 / 6,573 (0.11x)     &88,557 / 22,159 (0.25x)  &173,029 / 62,203 (0.36x)   \\
tcpdump   &14,463 / 99,652 (6.89x)   &28,762 / 204,787 (7.12x)  &80,374 / 613,256 (7.63x) &102,871 / 823,313 (8.00x)   \\
\midrule
Average increase &1.60x           &2.46x                  &3.48x                 &3.70x  \\
   \bottomrule
    \end{tabular}  
    \begin{tablenotes}  
\item[*] The value in bracket represents the corresponding increment of UltraFuzz compared to AFL.
\end{tablenotes} 
\end{table*}

We also compare AFL and UltraFuzz using the number of fuzzing tasks completed. The experiments were conducted with 8, 16, 32, and 64 cores, running for 24 hours, respectively. 
As Table \ref{table:task} shows, in general, UltraFuzz outperforms AFL on most (46/64) of the cases, with the highest increment of 11.40x (tifftobw).
It is worth noting that, with the number of cores increasing, the averaged task increments are getting large accordingly, ranging from 1.60x to 3.70x.
We can conclude from the result that UltraFuzz has a higher efficiency than AFL because UltraFuzz can complete more tasks than AFL at the same time.
A reason behind this is that, in the single-core fuzzing process of AFL, different fuzzing instances can repeatedly get blocked by the same slow inputs,
while in UltraFuzz, such low-quality seeds are removed immediately before affecting other instances.
More importantly, the efficiency advantage of UltraFuzz is getting obvious with the increase of the core. 
This is because based on the efficient seed and coverage information synchronization, UltraFuzz can execute more high-quality and fast seeds. 
In summary, UltraFuzz outperforms AFL in terms of completed tasks, and the super-linear acceleration is also reflected in the increments.

\subsubsection{Explanation}
\textbf{Explanation 1: Global energy scheduling optimization.}
According to AFL’s design, when it tests a seed with the random strategy and finds a new path, the energy (i.e., the number of mutation chances) assigned to this seed will be doubled. Though UltraFuzz does not optimize the scheduling algorithm at the code level (i.e., the energy assignation and mutation strategy of UltraFuzz are the same as those in AFL), the parallel mechanism of UltraFuzz allows it to test different seeds at the same time. Since these seeds are selected and marked as favored by the scheduler when UltraFuzz finds new paths during fuzzing these seeds, the energy assigned to these seeds will be doubled, which can promote the generation of more test cases.
For example, for favored seeds $s_1$ and $s_2$, assigning energy $e_1$ and $e_2$ to them, they may discover the same path $p_3$ with generating test case $s_3$. AFL-single selects seeds in order, by fuzzing $s_2$ after $s_1$. Once $s_1$ has discovered path $p_3$ and doubled the energy, when $s_2$ discovers path $p_3$ again, the energy is not doubled again. So, the total energy assigned to these two seeds in AFL-single is $(e_1*2 + e_2)$. In contrast, UltraFuzz can test these two seeds concurrently. When UltraFuzz mutates $s_1$ and $s_2$ in different fuzzing instances to generate test cases and both discover path $p_3$, the energy allocated to both $s_1$ and $s_2$ will be doubled. Thus, the total energy assigned to these two seeds in UltraFuzz is $(e_1*2+e_2*2)$, which is more than that in AFL-single. In other words, \textit{using the parallel mechanism, UltraFuzz can optimize global energy scheduling by giving favored seeds more energy.}

To verify this explanation, we analyzed the distribution of energy allocation in UltraFuzz and AFL-single during fuzzing \emph{readelf} with 32 units of computing resources. We recorded the energy assigned to each seed and the times of doubling energy for each seed during the random strategy. Fig. \ref{fig:energy} shows the energy distribution and the doubling energy times of UltraFuzz and AFL-single. From Fig. \ref{fig:energy}(a), we can see that both AFL-single and UltraFuzz allocate much energy to seeds whose serial numbers are closer to 0.
The reason is that these seeds are regarded as favored and their performance scores are higher than other seeds, which means these seeds are more likely to discover new paths and branches. For these high-quality seeds, particularly those with serial numbers less than 500, UltraFuzz allocates much more energy to them than AFL-single, which improves the efficiency of energy utilization and makes UltraFuzz reach a higher path coverage than AFL-single in \emph{readelf}. 
Moreover, from Fig. \ref{fig:energy}(b), the times of doubling the energy when fuzzing \emph{readelf} for many seeds in UltraFuzz were more than those in AFL-single, particularly for the seeds whose serial numbers are less than 100. The total doubling times for UltraFuzz were 3$\times$ greater than those of AFL-single, which were 1,501 and 544, respectively. This is consistent with our explanation. This demonstrates that UltraFuzz assigns more energy to seeds with a high probability of finding new paths and branches than AFL-single.

\begin{figure}[h]
\setlength{\abovecaptionskip}{0cm}
  \centering
  \includegraphics[width=0.9\linewidth]{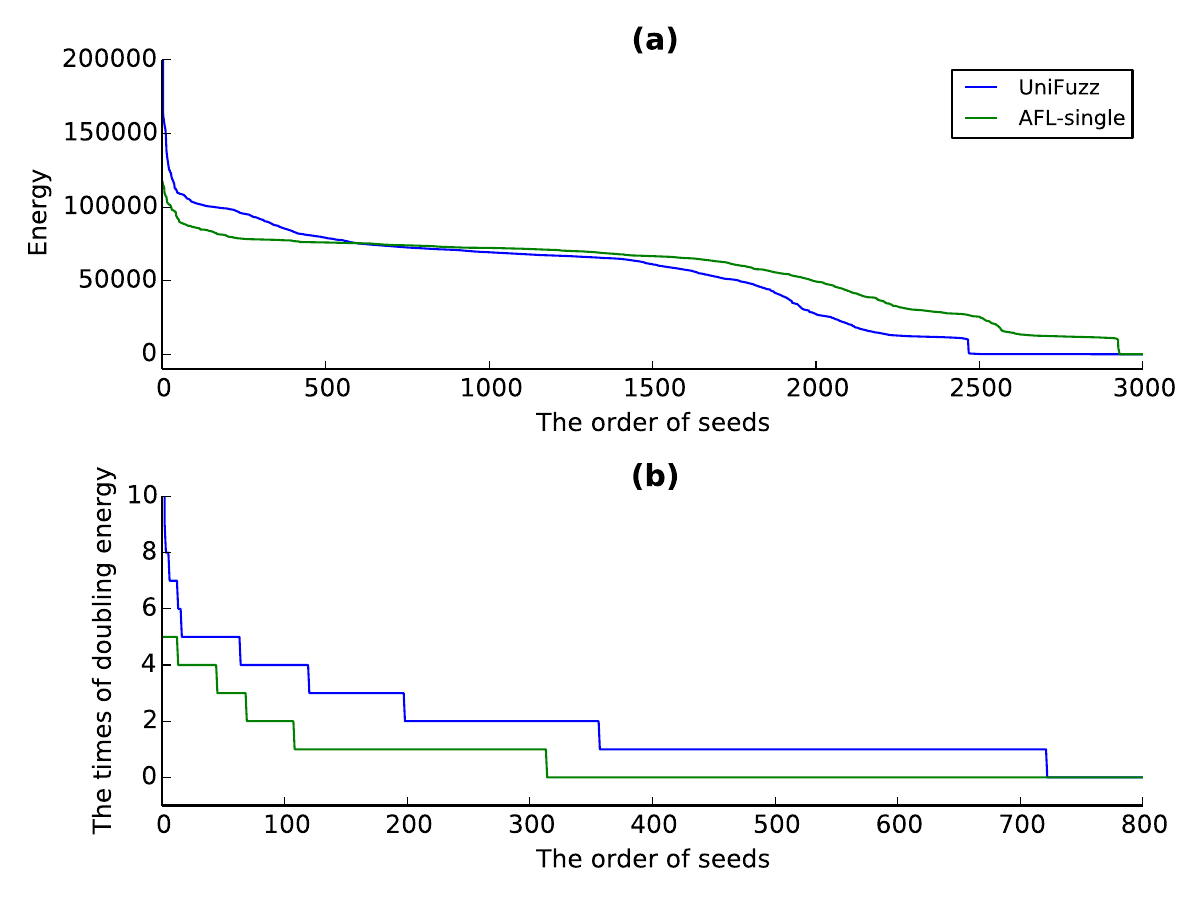}
  \caption{The distribution of energy assigned to each seed and the times of doubling energy in fuzzing \emph{readelf}. Since lots of seeds are not fuzzed (i.e., the energy assigned to these seeds is 0), we only focus on the seeds that have been fuzzed and sort them by the energy assigned to them. }
  \label{fig:energy}
\end{figure}

To further validate how much such global energy optimization affects the result, we close the double energy scheme of UltraFuzz and ran an additional experiment. Results show that 
5.3\% (513/9,707) patch coverage and 4.2\% (357/8,514) branch coverage are reduced. This implies that global energy optimization is only part of the reason for super-linear acceleration. 
In summary, parallel mechanisms can help optimize AFL's energy scheduling from a global perspective, but not the only reason.

\textbf{Explanation 2: Escape the local optimal seed.}
Most of the feedback-driven fuzzers are based on ad-hoc algorithms. The fuzzing process can start from a random initial seed and use a search-based strategy to alternatively approach the optimal solution. However, this procedure inevitably gets stuck in a local optimal seed and repeats redundant work. In serial fuzzing, we chose the best seed each time, which is similar to a depth-first search strategy. However, for parallel fuzzing with task distribution, different fuzzing instances can fuzz different seeds simultaneously, which is like a width-first search strategy. In such a parallel situation, a fuzzing instance that jumps out of the local optimal seed can notify the rest by instantly synchronizing the fuzzing status. In this way, fuzzing instances can escape local optimal seeds more easily to explore more seeds.

\begin{figure}[h]
\setlength{\abovecaptionskip}{-0.5cm}
  \centering
  \includegraphics[width=\linewidth]{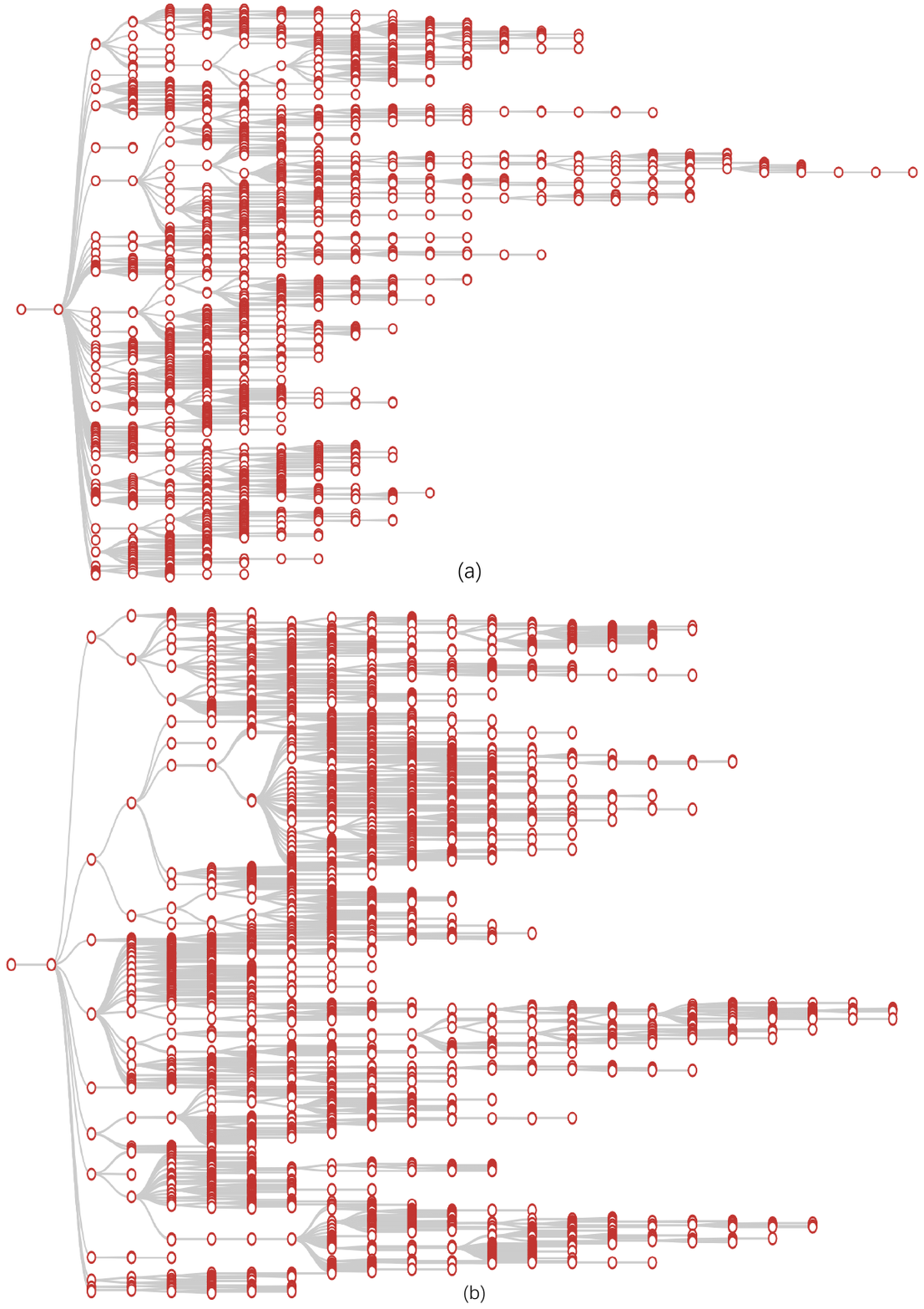}
  \caption{The relationship of seeds generated by AFL and UltraFuzz fuzzing objdump under 128 units computing resource.}
  \label{fig:activeseed}
\end{figure}

To verify this explanation, we construct an inherit-relation graph of the seeds generated during a fuzzing campaign. As an example, Fig. \ref{fig:activeseed} (a) and (b) show the relation of seeds generated during fuzzing \texttt{objdump} by AFL and UltraFuzz, respectively.  In the graph, each node indicates a seed, and the line between the two nodes indicates the inherited relationship. The campaign starts from the initial seeds at the far left of the graph. To save space, we only keep the seeds that discover new seeds in the graph, called \textit{active seeds}, and omit the rest (i.e., the leaf nodes). From a view of macroscopic, the seeds of UltraFuzz are distributed in a broader range than that of AFL. We can observe that UltraFuzz discovered more paths than AFL, and most of the paths are explored at a deeper level. Quantitatively, AFL generates 5,828 seeds in total, among which 1,450 are active, accounting for 24.9\%. For UltraFuzz, 2,183 out of 8,335 are active seeds, which accounts for 26.2\%. Therefore, we can conclude that UltraFuzz can generate more active seeds that can discover new paths compared to AFL, which is consistent with our explanation that UltraFuzz can escape the local optimum trap and explore more paths. 
In summary, instant synchronization in parallel fuzzing can help to fuzz instances get out of the local optimal seed faster and improve the performance,  but it is not the only reason, either.



These two Explanations can also explain why UltraFuzz can produce more test cases than AFL-single in Table \ref{table:resource}. \textbf{Reason 1}: AFL-based fuzzers prioritize seeds that are short in size and execute fast. According to Explanation 1, high-quality seeds (i.e., seeds that execute fast) will be preferred and given more energy by UltraFuzz once they discover new paths. Thus, UltraFuzz will have more fast seeds mutated, and consequently, produce more test cases (which are also fast) than AFL in the given time. \textbf{Reason 2}: AFL-single with virtual scaling might get stuck when a seed is very slow, while UltraFuzz with physical scaling will only have one instance get stuck, the other can proceed as usual. This reason is similar to Explanation 2.

\subsection{Exposed Vulnerabilities}

\begin{table}
  \centering
  \scriptsize
  \caption{The vulnerabilities found by these five tools}
  \label{table:vul}
\begin{threeparttable}
\begin{tabular}{p{1cm}|p{2.25cm}|p{2.7cm}|p{0.25cm}|p{0.5cm}}
\toprule
Program   & Type                  & Position &St. &Tool\\
\midrule
boringssl & double-free& asn1\_lib.c:460&N&U \\
boringssl & SEGV &asn1\_lib.c:459 &N&All\\
size & free invalid pointer &coffgen.c:1782&F&All\\
size &heap-use-after-free &elf.c:2604&F &U\\
size&SEGV&opncls.c:978&F&U\\
objdump& SEGV &elf.c:358&S&U\\
infotocap &SEGV &strchr.S:88&S&All\\
infotocap& heap-buffer-overflow&vg\_replace\_malloc.c:307&S&U\\
gnuplot&SEGV&plot2d.c:2218&W&All\\
gnuplot&heap-buffer-overflow&cgm.trm:1014&F &UA\\
gnuplot&SEGV&plot2d.c:1721&S&A\\
gnuplot&SEGV&graph3d.c:1962 &S&A\\
gnuplot&SEGV& hidden3d.c:1639&S&A\\
gnuplot&SEGV& iofclose.c:53 &S&A\\
gnuplot&SEGV&plot2d.c:3464&F &UA\\
gnuplot&heap-buffer-overflow&vg\_replace\_strmem.c:459&F &U\\
gnuplot&heap-buffer-overflow&datafile.c:2960&F &U\\
gnuplot&heap-use-after-free&misc.c:250&F&U\\
gnuplot&heap-use-after-free&vg\_replace\_strmem.c:1644&F&U\\
gnuplot&global-buffer-overflow&set.c:5184&W&U\\
gnuplot&double-free&graph3d.c:752&W&U\\
gnuplot&SEGV&tkcanvas.trm:1474 &W&U\\
gnuplot&SEGV&graphics.c:2333 &W&U\\
gnuplot&SEGV&axis.c:2412 &W&U\\

\bottomrule
\end{tabular}
\begin{tablenotes}  
\item[*] In the st. column, ``N'' indicates the vulnerability has never been reported before. ``F'' means the vulnerability has been fixed. ``W'' means the vendor wouldn't fix or accepted it. ``S'' means the vulnerability has been submitted to the vendor. In the Tool column, ``U'', ``A'' stand for UltraFuzz, AFL-single, and ``All'' means found by all the tools.
\end{tablenotes} 
\end{threeparttable}
\end{table}

As listed in Table \ref{table:vul}, these tools detected 24 vulnerabilities in total. UltraFuzz found 20 of them, more than the other five tools. In Google fuzzer-test-suite, UltraFuzz found two vulnerabilities in \emph{boringssl}, which were not reported by Google fuzzer-test-suite before. Particularly, the double-free vulnerability in \emph{boringssl} is only detected by UltraFuzz. In addition to Google fuzzer-test-suite, these tools detected 22 vulnerabilities in other real-world programs in total. Among them, UltraFuzz discovered 18 vulnerabilities. UltraFuzz finds three vulnerabilities in \emph{size}. One is a heap-use-after-free vulnerability, which is triggered by the \textit{bfd\_section\_from\_shdr} function. One is triggered by the \textit{\_bfd\_coff\_free\_symbols} function when attempting to free an invalid pointer. The other is a segmentation fault in \textit{\_objalloc\_alloc}, which was only detected by UltraFuzz. We have submitted these vulnerabilities to their vendors and the three vulnerabilities found in \emph{size} have been acknowledged. 
For \emph{gnuplot}, UltraFuzz found twelve vulnerabilities and six of them have been acknowledged, with detecting nine vulnerabilities. Compared to UltraFuzz, AFL detected seven vulnerabilities, less than UltraFuzz. Finally, the vulnerabilities found by UltraFuzz, AFL, PAFL, EnFuzz, and AFL-P in these programs are 20, 10, 4, 4, and 4, respectively. In conclusion, UltraFuzz outperforms other tools in finding vulnerabilities.

\subsection{Performance Overhead}

\subsubsection{Comparison with AFL-P in 128 units computing resources}

\begin{table}
  \centering  \scriptsize
  \caption{Overhead of UltraFuzz and AFL-P in 128 units computing resources}
  \label{table:overhead}
    \begin{tabular}{l|cccc}
   \toprule
      \multirow{2}{*}{{Program}}& \multicolumn{2}{c}{{UltraFuzz}} & \multicolumn{2}{c}{{AFL-P}}       \\ \cline{2-5} 
&Time(s)&Percent&Time(s)&Percent \\
   \midrule
boringssl&10,178&2.21\%&519&0.11\%\\
freetype&3,808&0.83\%&2,307&0.5\%\\
libcxx&8,650&1.88\%&2,875&0.62\%\\
libxml&4,298&0.93\%&1,428&0.31\%\\
re2&18,609&4.04\%&3,669&0.8\%\\
libarch&4,825&1.05\%&1,387&0.3\%\\
size&4,105&0.89\%&2,535&0.55\%\\
readelf&3,775&0.82\%&26,783&5.81\%\\
objdump&3,806&0.83\%&20,514&4.45\%\\
avconv&3,788&0.82\%&3,163&0.69\%\\
infotocap&4,903&1.06\%&1,882&0.41\%\\
pdftotext&3,800&0.82\%&12,428&2.7\%\\
tiff2bw&3,898&0.85\%&18,973&4.12\%\\
ffmpeg&4,304&0.93\%&2,190&0.48\%\\
gnuplot&4,953&1.07\%&2,153&0.47\%\\
tcpdump&5,573&1.21\%&2,072&0.45\%\\
\midrule
Average&5,829&1.26\%&6,555&1.42\%\\
   \bottomrule

    \end{tabular}  
    \vspace{-0.4cm}
\end{table}

To calculate the performance overhead of UltraFuzz, we classify the states of each working core into the \textit{fuzzing state} and the \textit{non-fuzzing state}. In the fuzzing state, the core is conducting a fuzzing task, while in the non-fuzzing state, the core may be requesting a new task, evaluating new seeds, communicating with the scheduler, or synchronizing with the database. 
Thus, we define the performance overhead of the system as \textit{the summation of time durations that the working cores are non-fuzzing}. We use $t$ and $n$ to represent the testing time and working core number, respectively. Then, we use $t_{i}$ to represent the time duration that the $ i^{th}$ core is on the fuzzing state. It is easy and accurate to measure the time that a core is undertaking a fuzzing task. On this basis, the global performance overhead of UltraFuzz can be calculated as $O = (1- \frac{\sum_{1} ^{n-1} t_{i}}{tn}) \times 100\%$. Notably, the scheduler is always in a non-fuzzing state. Thus, the core that the scheduler resides incurs a constant overhead, which should be included. We only measure the fuzzing state time of the rest $n-1$ cores of UltraFuzz.

Since the synchronization overhead increases with the parallel scale, we use an experiment configuration of 128 units of computing resources to test the maximum overhead available. Thus, the overall testing time is 128*3600 seconds. Then, the temporal summation of all the 127 working cores in the fuzzing state was recorded to calculate the overhead. As Table \ref{table:overhead} shows, the overhead of UltraFuzz on 9/16 programs is under 1\%. On average, the overhead is 1.26\%, which is acceptable for a parallel fuzzer with 128 cores. As a comparison, we also record the overhead of AFL-P. AFL-P does not synchronize much information and does not support inter-machine mode. As a result, UltraFuzz (1.26\%) has a lower overhead than AFL-P (1.42\%) with 128 units of computing resources.
We infer the low overhead also contributes to the super-linear performance acceleration we have discussed in Section \ref{super}.

\begin{table}
  \centering \scriptsize
  \caption{Average overhead of UltraFuzz}
  \label{table:average overhead}
    \begin{tabular}{l|ccccc}
   \toprule
   Overhead  &8 units &16 units &32 units &64 units &128 units \\
   \midrule
Time(s)&3,611&3,634&3,668&3,726&5,828\\
Percentage&12.54\%&6.31\%&3.18\%&1.62\%&1.26\%\\
   \bottomrule
    \end{tabular}  
\end{table}

Intuitively, the performance overhead should increase as the parallel scale increases. This is because inter-machine communication can aggravate the system overhead. However, in reality, as Table \ref{table:average overhead} shows, the performance overhead of UltraFuzz decreases as the parallel scale increases. This is explainable because, according to the way we calculate the overhead, the core that the scheduler occupies is always regarded as being at a non-fuzzing state, incurring overhead. The more cores we use, the lower percentage of overhead this core accounts for. 



\subsubsection{Comparison with AFL on 64 cores for 24 hours}

\begin{table*}
  \centering  \scriptsize
  \caption{Overhead of UltraFuzz and AFL with 64 cores running for 24 hours}
  \label{table:overhead}
    \begin{tabular}{l|rrrr|rrrrr}
   \toprule
      \multirow{2}{*}{{Program}}& \multicolumn{3}{c}{{AFL}} & \multicolumn{6}{c}{{UltraFuzz}}       \\ 
      \cline{2-10} 
       &Scheduling &Evaluation  &Fuzzing     &Evaluation &Seeds  &Update   &Socket &Database &Fuzzing\\
   \midrule
boringssl   &555    &38,288   &5,490,526     &23     &15,753    &48     &837    &13,533   &5,499,192\\
freetype    &1,191   &73,980   &5,454,320    &790    &17,324    &1,282  &170    &41,343   &5,470,617 \\
libcxx      &2,567   &38,778   &5,488,106    &226    &48,642    &374    &318    &41,889   &5,438,803\\
libxml      &1,472   &75,585   &5,452,065    &191    &2,987     &231    &112    &4,946    &5,521,052\\
re2         &1,793   &54,871   &5,472,745    &153    &116,437   &384    &1,879  &79,849   &5,331,408 \\
libarch     &986    &37,567   &5,490,919     &140    &4,126     &181    &63     &5,574    &5,519,755\\
size        &739    &38,596   &5,490,126     &106    &23,862    &89     &361    &18,841   &5,486,125\\
readelf     &1,957   &48,394   &5,479,183    &434    &14,056    &776    &167    &29,425   &5,486,083\\
objdump     &1,160   &43,599   &5,484,775    &256    &4,020     &206    &46     &6,797   &5,518,046\\
avconv      &1,957   &69,802   &5,457,774    &5,882   &7,963      &3,429   &130    &45,422  &5,479,453\\
infotocap   &458    &31,347   &5,497,796     &1,701  &2,483     &154    &11     &1,072   &5,527,703\\
pdftotext   &1,412   &22,984   &5,505,113    &264    &1,731     &112    &73     &1,416   &5,526,205\\
tiff2bw     &1,179   &52,808   &5,475,526    &119    &13,593    &121    &211    &9,613   &5,507,194 \\
ffmpeg      &538    &41,596   &5,487,412     &4,896  &8,613     &3,205  &167    &72,308  &5,451,054 \\
gnuplot     &4,828   &62,942   &5,461,823    &3,029  &15,070    &2,609  &46     &92,810   &5,352,737\\
tcpdump     &3,008   &37,336   &5,489,098    &381    &72,426    &972    &271    &80,733   &5,376,202\\
\midrule
Average percentage     &0.03\%  &0.87\%   &99.10\%     &0.02\%  &0.42\%   &0.02\%  &0.01\%  &0.62\%  &98.89\%\\
   \bottomrule

    \end{tabular}  
\end{table*}
In this section, we measure the time spent in AFL and UltraFuzz in fine granularity. 
We run both fuzzers with 64 cores for 24 hours. The total testing time is 64 cores * 24h * 3600s = 5,529,600s.
For AFL, we measure the time spent on scheduling, evaluation, and fuzzing, respectively, which are listed as column 2 - column 4 in Table \ref{table:overhead}.
For UltraFuzz, in addition to evaluation (column 5) and fuzzing (column 10), we further distinguish the time spent on downloading and uploading seeds (column 6), 
update seeds and bitmap (column 7), inter-machine socket communication (column 8), and database-based synchronization (column 9).
The last row of Table \ref{table:overhead} represents the average percentage of the time spent in each category.
The results can provide a fine-grained resolution of the overhead of AFL and UltraFuzz.
We can see that both AFL and UltraFuzz spent around 99\% of the testing time on fuzzing, which means the overheads are both around 1\%. 
More specifically, compared to AFL, the overhead increment of UltraFuzz is around 0.2\% (i.e., 1.09\% - 0.90\%), which is acceptable. 
Among the overhead of UltraFuzz, seed transfer (0.42\%) and data-based synchronization (0.62\%) consume the most time. 
This is consistent with our intuition as the seed is relatively large compared with other fuzzing information.
Notably, the inter-machine communication via socket incurs very little overhead, which is less than 0.01\%. 
In summary, UltraFuzz has high efficiency, and its overhead is close to that of AFL.

\subsection{Effectiveness of Seed Caching}

\begin{table}
  \centering  \scriptsize
  \caption{Effectiveness of Seed Caching}
  \label{table:caching}
    \begin{tabular}{l|rrr}
   \toprule
      Program &Hit ratio    &Caching enabled  &Cache disabled\\
   \midrule
boringssl   &0.993   &15,689   &78,425 (5.00x)    \\
freetype    &0.444   &15,187   &87,182 (5.74x)  \\
libcxx      &0.799   &48,223   &110,458 (2.29x)  \\
libxml      &0.294   &2,515    &3,999 (1.59x)  \\
re2         &0.994   &113,827  &1,326,020 (11.65x)  \\
libarch     &0.580   &3,734    &16,278 (4.36x)   \\
size        &0.975   &23,722   &305,729 (12.89x)   \\
readelf     &0.545   &13,190   &197,976 (15.01x)  \\
objdump     &0.700   &3,587    &59,675 (16.63x)  \\
avconv      &0.145   &437      &12,710 (29.08x)  \\
infotocap   &0.420   &689      &21,616 (31.37x)   \\
pdftotext   &0.859   &1,399    &21,847 (15.62x)  \\
tiff2bw     &0.989   &12,262   &569,194 (46.42x)  \\
ffmpeg      &0.105   &920      &14,628 (15.90x)   \\
gnuplot     &0.199   &11,838   &133,095 (11.24x)  \\
tcpdump     &0.834   &71,964   &586,704 (8.15x)  \\
\midrule
Average     &0.617   &-        &14.56x\\
   \bottomrule

    \end{tabular}  
\end{table}
To prove the effectiveness of the seed caching mechanism in UltraFuzz, as Table \ref{table:caching} shows, we recorded the cache hit ratio (column 2) and the time consumed in downloading seeds when seed caching is enabled (column 3)/disabled (column 4). The numbers in the brackets represent the corresponding increments of seed downloading time when seed caching is disabled compared to enabled.
The experiments were conducted with 64 cores running for 24 hours.
The hit ratios range from 0.105 to 0.993 with an average value of 0.617, which demonstrates that the seed caching is meaningful. 
The seed downloading time increments range from 1.59x to 46.42x with an average value of 14.56x, which demonstrates the effectiveness of seed caching. 
It is worth noting that a high hit ratio (e.g., boringssl) does not necessarily lead to a very high increment, and a low hit ratio (e.g., ffmpeg and avconv) can also bring significant increments. 
In summary, the seed caching mechanism is effective in shortening the time consumed in downloading seeds, but its performance is also affected by the characteristic of the seeds, e.g., the seed size.

\subsection{Effectiveness of the Elasticity Mechanism}
\begin{table}
  \centering \scriptsize
  \caption{Effectiveness of Elasticity Mechanism .}
  \label{table:elastic}
    \begin{tabular}{l|rr}
   \toprule
   \multirow{2}{*}{{Program}} & \multicolumn{1}{c}{{Number of Fuzzing Tasks}}  & \multicolumn{1}{c}{{Evaluation Time (s) }}     \\ 
   \cline{2-3} 
            & {Enable / Disable Elasticity} & {Enable / Disable Elasticity}  \\
   \midrule
boringssl &2,102,387 / 1,125,994 (0.54x)     & 23 / 24 (1.04x)                  \\
freetype  &119,622 / 97,077 (0.81x)       & 790 / 722 (0.91x)                   \\
libcxx    &863,904 / 748,660 (0.87x)       & 226 / 319 (1.41x)                  \\
libxml    &61,297 / 39,828 (0.65x)          & 191 / 242 (1.27x)                   \\
re2       &5,198,349 / 5,669,606 (1.09x)     & 153 / 152 (0.99x)                  \\
libarch   &89,128 / 59,501 (0.67x)          & 140 / 119 (0.85x)                   \\
size      &1,054,117 / 998,718 (0.95x)     & 106 / 109 (1.03x)                  \\
readelf   &137,259 / 87,767 (0.64x)       & 434 / 416 (0.96x)                   \\
objdump   &78,909 /87,140 (1.10x)        & 256 / 260 (1.02x)                  \\
avconv    &2,167 / 2,097 (0.97x)         & 5,882 / 6,047 (1.03x)                    \\
infotocap &7,000 / 6,836 (0.98x)         & 1,701 / 3,294 (1.94x)                   \\
pdftotext &153,019 / 135,092 (0.88x)       & 264 / 294 (1.11x)                   \\
tiff2bw   &505,215 / 450,350 (0.89x)       & 119 / 108 (0.91x)                  \\
ffmpeg    &5,093 / 4,132 (0.81x)         & 4,896 / 5100 (1.04x)                   \\
gnuplot   &62,203 /37,986 (0.61x)        & 3,029 / 3,005 (0.99x)                 \\
tcpdump   &823,313 / 622,250 (0.76x)       & 381 / 385 (1.01x)                  \\
\midrule                      
Average increase   &0.83x           &1.09x                 \\
   \bottomrule
    \end{tabular}  
    \begin{tablenotes}  
\item[*] The value in bracket represents the corresponding increment of UltraFuzz when elasticity mechanism is disabled compared to enabled.
\end{tablenotes} 
\end{table}

To prove the effectiveness of the elasticity mechanism of evaluating instances in UltraFuzz, 
we use the number of completed fuzzing tasks and the evaluation time as metrics. 
By enabling and disabling the elasticity mechanism, we can show the effectiveness of the elasticity mechanism via the two metrics.
The experiments were conducted with 64 cores running for 24 hours. 
As Table \ref{table:elastic} shows, when the elasticity mechanism is disabled, most of the fuzzing tasks completed (14/16) are reduced, resulting in 0.54 - 0.97 to that when elasticity is enabled. 
The average decrement of the completed fuzzing task is 17\%.
As for the evaluation time, most of the cases (10/16) are increased, from 1.01x to 1.41x, with an average increment of 9\%. 
However, the results also show that disabling the elasticity mechanism does not necessarily result in a decrement in the number of completed fuzzing tasks nor an increment in the evaluation time.
A small portion of the results is opposed. This is because our algorithm for balancing the number of fuzzing instances and evaluating instances is heuristical. 
We cannot guarantee to have the optimal arrangement for all the programs under test. 
Nevertheless, the results prove that the elasticity mechanism is useful in most cases in improving the efficiency of UltraFuzz.

\section{Discussion}

Parallel fuzzing in a distributed environment amplifies the resource-wasting problem caused by the random nature of fuzzing. With centralized dynamic scheduling, UltraFuzz can optimize seed selection, energy scheduling, fuzzing status synchronization, and task dispatching over all the working nodes from a global perspective. Based on the techniques including on-demand task dispatching, hierarchical information synchronization, and elastic seed evaluation computing power allocation, UltraFuzz can filter duplicate (redundant) seeds and avoid task conflicts, achieve instant synchronization, balance workload, elastically allocate computing power, and be compatible with the dynamic change of computing capability. All these benefit resource utilization and improve performance.

\textbf{Super-linear acceleration.}
In the experiments, we had an empirical observation that UltraFuzz on n cores for 1 hour performs better than AFL-single on 1 core for n hours. We call it 
super-linear acceleration. We believe this result is caused by superimposing reasons.
Except for the reason that resource utilization efficiency is improved in UltraFuzz, we also propose another two explanations and validate them with experiments. 
As we have discussed in Section \ref{super}, one is that UltraFuzz can optimize AFL's energy scheduling from a global perspective by assigning more energy to seeds with a high probability of finding new paths.
Another explanation is that the instant synchronization scheme in UltraFuzz helps the fuzzing instances jump out of the local optimal seed fast to discover and explore better seeds.
Experiments show that super-linear acceleration does not exist with classical AFL synchronization (i.e., AFL-P, see Table \ref{table:resource3} in the Appendix), or independently running instances. This is because a necessary condition of super-linear acceleration is the instant fuzzing status synchronization among instances. Without fast and efficient synchronization, seeds cannot be de-duplicated timely, and fuzzing tasks cannot be dispatched reasonably, as a result, task conflicts and redundant executions would greatly deduct the system efficiency.
We also exclude the possibility that special inputs might block or slow the execution of AFL-single.
We use the same initial inputs for all the tools we compare. If special inputs slow AFL-single, then all the tools should outperform AFL-single. However, only UltraFuzz has super-linear acceleration manifested, and the rest tools perform less well as AFL-single.

However, we don't think super-linear acceleration is persistent.
In a fuzzing process, it is obvious that the coverage increases actively at the beginning and tends to smooth when the program has been fuzzed for a long time enough. 
This is because, for any program under test, since the program size is fixed, a fuzzer is difficult to find new coverage when a majority of the program has been explored. 
To simplify this process, we roughly divide the whole fuzzing process into two phases: the \textit{active phase} and the \textit{smooth phase}, though without a clear boundary.
Most of the optimization techniques are clearly effective in the active phase. Once fuzzing enters the smooth phase, the test tends to be saturated as most of the paths have been explored and it is difficult to see an obvious difference by comparing coverage. Thus, we think that super-linear acceleration only manifests obviously at the active phase. Therefore, when setting the experiment, we always follow two criteria:
1) select relatively large-size programs to extend the active phase; 2) limit the computing resources (cores and time) to restrict the test within the active phase. 
Increasing the testing resource (time or core) is definitely useful to discover vulnerabilities, however,  
blindly increasing the testing resource might conceal the difference when comparing the performance of different tools.
Though UltraFuzz runs tests only for one hour in our experiment, it leverages a large number of cores to share the task and shorten the time. The workload is divided and dispatched to different instances to fuzz concurrently. Essentially, parallel fuzzing trades cores for time. 
Modern fuzzing evaluation follows the guidance proposed by Klees \textit{et al.} \cite{klees2018evaluating} that a fuzzing campaign should at least take 24 hours. However, this claim is for single-core fuzzing. For parallel fuzzing, the real testing time should multiply the number of cores to measure the capacity of work.
Besides, we have AFL running tests long enough as the baseline to validate the tests.
 As for the potential randomness of fuzzing, we choose to reduce such impact by repeating the experiment instead of lengthening the test time. We also conduct a statistical significance analysis for the experiment result.

Though UltraFuzz has high resource utilization efficiency,  itself (like all other studied fuzzers) is still subject to Boehme's ``empirical law of exponential cost". However, this study can provide insight and direction to future research on parallel fuzzing. 


\textbf{Advantages.}
Beyond the technical advantages over the state-of-the-art baselines, UltraFuzz can be useful in promoting other variants of fuzzing, such as differential fuzzing \cite{petsios2017nezha,nilizadeh2019diffuzz,noller2020hydiff,noller2021qfuzz} and directed fuzzing \cite{bohme2017directed, chen2018hawkeye, liang2019sequence, nguyen2020binary, wang2020sok}. 
Differential fuzzing detects bugs by providing the same input to different implementations of the same application and observing differences in their execution. Differential testing is well-suited to find semantic or logic bugs that do not exhibit explicit erroneous behaviors like crashes or assertion failures, which complements traditional fuzz testing. UltraFuzz can improve the efficiency of differential fuzzing by running the executions in parallel. 
Directed fuzzing focuses on target locations (e.g., the bug-prone zone) and spends most of its time budget on reaching these locations without wasting resources stressing unrelated parts. By giving more mutation chances to seeds that are closer to the target, directed fuzzing can reach the target locations gradually.
UltraFuzz can help to improve the efficiency of directed fuzzing by assigning different targets to different instances and running them in parallel.

\textbf{Risk Statement.}
A centralized architecture with a scheduler has a vulnerable risk. If a failure happens to the scheduler or it is compromised, the whole system will be unworkable.

\textbf{Future work}
UltraFuzz is implemented on top of vanilla AFL, and we focus on improving the performance of fuzzing by optimizing the parallel scheme, such as the centralized dynamic scheduling and hierarchical information-sharing. UltraFuzz can integrate with works from an orthogonal direction, such as improving execution speed~\cite{schumilo2017kafl,zhang2018ptfuzz,chen2019ptrix,libafl}, optimizing mutation strategy~\cite{chen2018hawkeye,lyu2019mopt,li2019v,you2019profuzzer} and power scheduling \cite{yue2020ecofuzz,bohme2017directed}, or improving the diversity
by taking advantage of different fuzzers \cite{enfuzz, guler2020cupid}. 
We leave these as future work.

\section{Threat to validation}
(1) The coverage sharing scheme of UltraFuzz might be affected by extreme situations, such as programs with randomization. In such a case, the same inputs and same program may not lead to the same execution trace and thus the bitmap coverage may be different, too.  To reduce such impact, we repeat each experiment \textbf{10} times. 

(2) In some experiments, for tools that do not support inter-machine mode, we virtually expand the experiment scale by extending the test time. However, such virtual scaling cannot reflect the exact performance in a distributed environment. In a virtual configuration, the communication cost among nodes is neglected. Notably, the coverage of virtual scaling is definitely increasing, however, we cannot guarantee coverage increase when scaling physically. Thus, virtual scaling by nature has advantages over physical scaling in comparison.

(3) In distributed fuzzing, the configuration of testing resources can affect performance.
Resource interference, involving CPU, memory, syscall, and file system, plays a significant role. 
We have tried to exclude such an effect in the experiments by leaving half of the cores unoccupied in each machine. However, unknown interference factors might still affect the performance.

(4) The characteristic of the program as well as the seed will also affect the results. For example, the seed size will greatly affect the performance of the caching scheme.

(5) The length of the fuzzing campaign might affect the results. We extend additional experiments to 24 hours, such as Section V.A (2), V.A (4), V.D (2), V.E, and V.F, to account for the potential difference at the beginning of the campaign.

\section{Related Work}

Most research improves fuzzing efficiency by designing novel algorithms and combining other techniques. AFLFast~\cite{aflfast} and Fairfuzz~\cite{fairfuzz} improve fuzzing efficiency by optimizing seed selection. The former tends to mutate seeds with low path frequency, and the latter modifies seeds whose hit count is relatively small. MobFuzz \cite{zhang2022mobfuzz} improves efficiency by multi-objective optimization. EcoFuzz \cite{yue2020ecofuzz} saves energy by optimizing the power scheduling of fuzzing. AFLGo~\cite{bohme2017directed} is directed to bug-prone locations to improve the probability of triggering bugs. Driller~\cite{driller}, and QSYM~\cite{qsym} leverage symbolic execution to improve fuzzing performance. VUZZER~\cite{vuzzer} and Angora~\cite{angora} use taint analysis to gather dynamic and static information from target programs to assist fuzzing. 
However, the performance of algorithm optimization is limited.


Fuzzing efficiency can also be improved by increasing computing resources to parallelize fuzzing tasks.
Early in 2010, Xie~\cite{grid} proposed a parallel framework, which leverages grid computing for large-scale fuzzing. The framework was implemented by storing fuzzing jobs in a server and scheduling remote clients to download these jobs. However, this kind of static scheduling results in an unbalanced partition of workload,  
and it only schedules fuzzing tasks to computing resources without trying to innovate the synchronizing and sharing mechanisms.
To parallelize AFL, AFL-P\cite{AFL1} extends its scalability by utilizing multiple processes running fuzzer instances to synchronize seeds between them. However, the scalability of AFL-P is limited because it is inaccessible when using computing resources across machines. Roving~\cite{roving} and disfuzz~\cite{disafl} addressed this problem with the help of the client/server structure. They share new seeds with each computing core within a fixed time interval, which enhances the scalability but produces redundant work and task conflicts, leading to a severe waste of computing resources. Li \textit{et al}.~\cite{large-scale} designed a tree structure to store coverage information as an alternative to bitmap. They leverage a polling mechanism to reduce redundant work and avoid conflicts, but this approach causes large performance consumption.

Several recent works focus on partitioning fuzzing tasks to avoid redundant work. In PAFL~\cite{PAFL}, local guiding information from each fuzzer instance is synchronized with global guiding information. According to the guiding information, PAFL assigns different task segments divided by grouping branches to different fuzzer instances. Though PAFL speeds up the fuzzing process, it cannot run a distributed system across multiple machines. Another work also called PAFL~\cite{PAFL-N}, collects dynamic execution information and dispatches parts of the target programs that have weak relationships, thus reducing redundant work. However, the disadvantage lies in the difficulty of accurately dividing the target program into parts. 
To address the scalability bottlenecks, Wen \textit{et al}.~\cite{Wen2017Designing} designed new operating primitives to speed up AFL. 
Although this work improves fuzzing performance with underutilized CPU cores, there is still room for improvement by scheduling fuzzing tasks and extending multi-core parallelism to distributed parallelism. P-fuzz~\cite{pfuzz1} leverages the computing resources of distributed systems to enhance fuzzing efficiency. It alleviates task conflicts in part by adopting database-centric architecture. But imbalances exist: some seeds are overused, while others are idle.

Different from the above works, EnFuzz~\cite{enfuzz} improves parallel fuzzing from an orthogonal direction.
It defines the diversity of fuzzers and chooses different fuzzers (AFL, AFLFast, FairFuzz, QSYM) to complement each other. Enfuzz’s main limitation is scalability. It only supports sharing guiding information based on the file system, and coverage information can differ greatly between fuzzers. Our approach improves fuzzing performance by optimizing the parallel scheme, and we can share various fuzzing information across machines.
LibAFL \cite{libafl} improves fuzzing performance by increasing the execution speed. Benefits from the optimization at compile-time and the implementation in Rust, it can keep runtime overhead minimal, which reaches 120k execs/sec in Frida-mode on a phone.
Differently, UltraFuzz focuses on optimizing the scheme of parallel fuzzing to save resources, which is an orthogonal direction. However, both works can be integrated.

\section{Conclusion}

This paper outlines the design and implementation of UltraFuzz, a fuzzing optimization toward resource-saving in a distributed environment. UltraFuzz globally optimizes seed selection, energy scheduling, inter-machine status synchronization, and task dispatch, and thus can overcome challenges such as task conflicts, workload imbalance, synchronization overhead, and dynamic change of computing core number. 
Experiments on real-world programs show that UltraFuzz outperforms state-of-the-art tools such as AFL, PAFL, and EnFuzz. 
We also discovered a counter-intuitive phenomenon that UltraFuzz can achieve super-linear acceleration compared to single-core AFL.
We proposed explanations of this phenomenon and validated them with additional experiments. 
Finally, 24 real-world vulnerabilities were also discovered. 

\section*{Acknowledgement}

The authors would like to sincerely thank all the reviewers for your time and expertise on this paper. Your insightful comments help us improve this work. This work is partially supported by the National University of Defense Technology Research Project (ZK20-17, ZK20-09), the National Natural Science Foundation China (62272472, 61902412, 61902416), and the HUNAN Province Natural Science Foundation (2021JJ40692, 2019JJ50729).



%

\bibliographystyle{unsrt} 
\bibliography{ref}

\clearpage

\appendix

\subsection{Path Coverage Comparison}
\label{path}

\begin{figure*}[ht]
\setlength{\abovecaptionskip}{0cm}
  \centering
  \includegraphics[width=\linewidth]{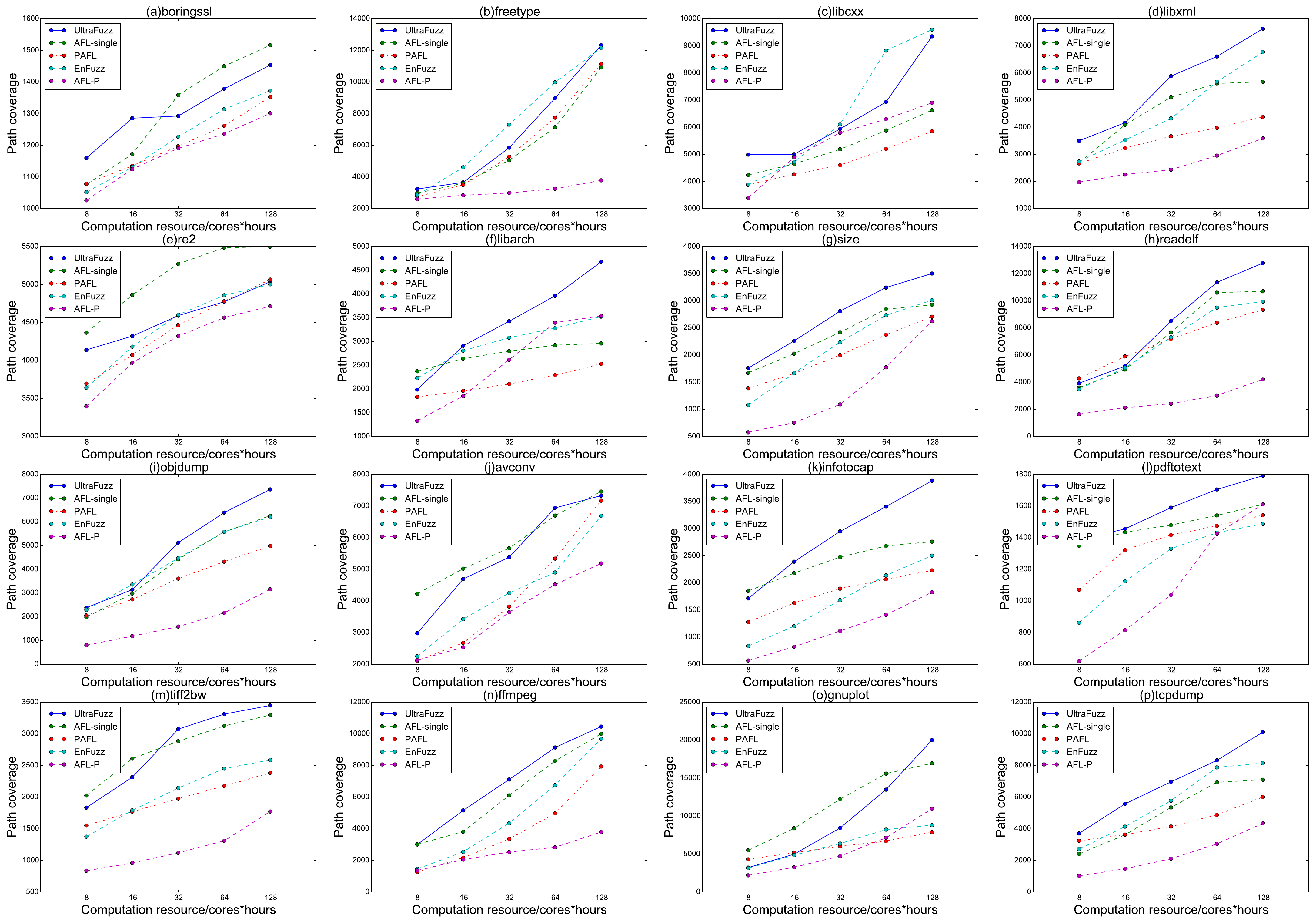}
  \caption{Comparison of path coverage reached by different tools with the same computing resources.}
  \label{fig:pathcoverage}
\end{figure*}

Fig.~\ref{fig:pathcoverage} plots the average number of paths discovered by these tools with different computing resources. We can see that the path coverage reached by each tool rises as computing resources increase on most programs.  Among these tools, UltraFuzz reaches the highest path coverage in most (13/16) programs, outperforming the other four tools. In particular, when compared with the baseline AFL-single,  UltraFuzz performs better on 13 programs with the same computing resources, such as \emph{freetype}, \emph{libcxx}, \emph{size} and \emph{readelf}. However, the other four tools do not reach a higher path coverage than that of AFL-single. 
In a word, the performance of path coverage is consistent with branch coverage.

In Table \ref{table:path coverage}, we also conduct a close comparison between UltraFuzz and AFL-single on path coverage. 
Particularly, we calculate the improved ratio of the path coverage reached by UltraFuzz and that reached by AFL-single. As Table \ref{table:path coverage} shows, UltraFuzz achieves higher path coverage than AFL-single on most programs from 8 units to 128 units. Moreover, as computing resources increase, the average speedup of UltraFuzz to AFL-single increases from 1.03x to 1.19x, which demonstrates that UltraFuzz also has super-linear performance acceleration, even more obvious. 
In summary, the results of path coverage are consistent with that of branch coverage.


%

\subsection{Path Cost for Each Program}
\label{appendix_path_cost}
    
Path cost is defined as the number of generated test cases divides the number of discovered paths in the campaign, which means the average number of test cases consumed on finding each path. Path cost can roughly reflect the quality of the test cases and the resource efficiency of the fuzzer.
We provide the comparison of different tools regarding path cost on different test programs in Fig. \ref{fig:total_average_energy}, which is consistent with the description in Section \ref{path_cost}. Among these tools, AFL-P has the highest path cost in most (13/16) programs. UltraFuzz achieves the lowest path cost in 6/16 of the programs, which is similar to the baseline AFL-single but better than all the other tools.

\begin{table*}[h]
\setlength{\abovecaptionskip}{0cm}
  \centering \scriptsize
  \caption{Path coverage increment of UltraFuzz compared to baseline with same computing resources.}
  \label{table:path coverage}
    \begin{tabular}{l|rrrrrr}
   \toprule
   Program  &8 units &16 units &32 units &64 units &128 units \\
   \midrule
boringssl&1,160(1.08x)&1,285(1.1x)&1,292(0.95x)&1,378(0.95x)&1,453(0.96x)\\
freetype&3,241(1.09x)&3,661(1.02x)&5,854(1.16x)&8,989(1.26x)&12,337(1.13x)\\
libcxx&4,993(1.18x)&5,005(1.08x)&5,941(1.14x)&6,933(1.18x)&9,354(1.41x)\\
libxml&3,501(1.3x)&4,170(1.02x)&5,886(1.15x)&6,611(1.18x)&7,635(1.34x)\\
re2&4,139(0.95x)&4,321(0.89x)&4,591(0.87x)&4,774(0.87x)&5,038(0.92x)\\
libarch&1,988(0.84x)&2,910(1.1x)&3,427(1.23x)&3,964(1.36x)&4,679(1.58x)\\
size&1,759(1.05x)&2,261(1.12x)&2,811(1.16x)&3,245(1.14x)&3,504(1.2x)\\
readelf&3,924(1.09x)&5,193(1.05x)&8,514(1.11x)&11,365(1.07x)&12,787(1.19x)\\
objdump&2,386(1.2x)&3,147(1.06x)&5,126(1.16x)&6,393(1.15x)&7,369(1.18x)\\
avconv&2,980(0.7x)&4,696(0.93x)&5,383(0.95x)&6,942(1.04x)&7,331(0.98x)\\
infotocaps&1,714(0.93x)&2,392(1.1x)&2,949(1.19x)&3,406(1.27x)&3,884(1.41x)\\
pdftotext&1,387(1.03x)&1,455(1.01x)&1,590(1.07x)&1,705(1.11x)&1,792(1.11x)\\
tiff2bw&1,834(0.91x)&2,316(0.89x)&3,076(1.07x)&3,313(1.06x)&3,449(1.05x)\\
ffmpeg&3,008(1.0x)&5,165(1.35x)&7,118(1.16x)&9,142(1.1x)&10,463(1.05x)\\
gnuplot&3,230(0.59x)&4,993(0.59x)&8,442(0.69x)&13,498(0.86x)&20,021(1.18x)\\
tcpdump&3,708(1.54x)&5,578(1.54x)&6,966(1.3x)&8,330(1.2x)&10,111(1.42x)\\
\midrule
Average increase&1.03x&1.05x&1.09x&1.11x&1.19x\\
   \bottomrule
    \end{tabular}  
\end{table*}

\begin{figure*}[ht]
\setlength{\abovecaptionskip}{0cm}
  \centering
  \includegraphics[width=\linewidth]{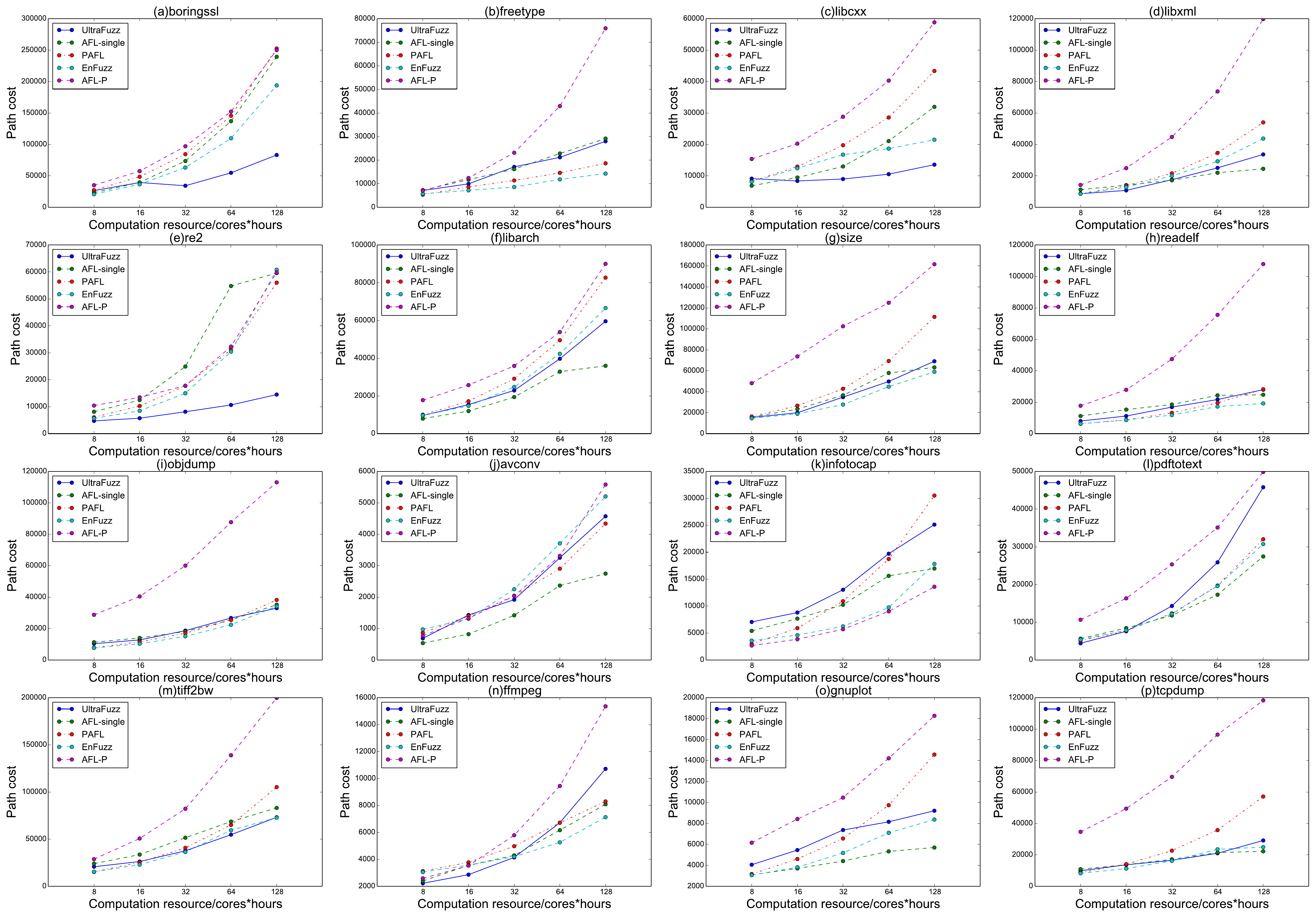}
  \caption{Comparison of different tools regarding path cost on different test programs.}
  \label{fig:total_average_energy}
\end{figure*}

\begin{table*}[t]
  \setlength{\abovecaptionskip}{0cm}
    \centering \scriptsize
    \caption{Branch coverage increment of AFL-P compared to baseline with same computing resources.}
    \label{table:resource3}
      \begin{tabular}{l|rrrrrr}
     \toprule
      Program  &8 units &16 units &32 units &64 units &128 units \\
     \midrule
  boringssl&2,417(0.97x)&2,436(0.98x)&2,473(0.96x)&2,492(0.95x)&2,517(0.95x)\\
  freetype&9,634(0.96x)&10,055(0.95x)&10,137(0.85x)&10,273(0.79x)&10,753(0.75x)\\
  libcxx&5,981(0.97x)&6,555(1.03x)&6,701(1.03x)&6,762(1.02x)&6,843(1.02x)\\
  libxml&5,629(0.92x)&5855(0.80x)&5980(0.76x)&6,263(0.77x)&6,571(0.80x)\\
  re2&5,647(0.92x)&6,029(0.97x)&6,068(0.97x)&6,090(0.97x)&6,096(0.97x)\\
  libarch&3,432(0.76x)&3,975(0.82x)&4,795(0.95x)&5,460(1.06x)&5,590(1.08x)\\
  size&2,258(0.64x)&2,498(0.67x)&3,002(0.76x)&3,772(0.92x)&4,063(0.98x)\\
  readelf&4,160(0.69x)&4,710(0.63x)&4,964(0.52x)&5,424(0.52x)&6,405(0.62x)\\
  objdump&3,366(0.65x)&3,854(0.62x)&4,629(0.65x)&5,212(0.69x)&6,116(0.78x)\\
  avconv&14,147(0.80x)&14,736(0.80x)&16,724(0.88x)&18,224(0.92x)&19,280(0.95x)\\
  infotocap&1,782(0.60x)&1,964(0.62x)&2,166(0.65x)&2,463(0.73x)&3,002(0.88x)\\
  pdftotext&1,738(0.94x)&1,789(0.97x)&1,830(0.99x)&1,860(1.00x)/&1,864(1.00x)\\
  tiff2bw&2,954(0.57x)&3,128(0.57x)&3,436(0.61x)&3,814(0.67x)&4,548(0.79x)\\
  ffmpeg&11,279(0.77x)&12,927(0.81x)&14,023(0.79x)&14,430(0.76x)&15,716(0.77x)\\
  gnuplot&9,013(0.74x)&10,394(0.66x)&11,818(0.59x)&13,556(0.58x)&17,441(0.72x)\\
  tcpdump&3,861(0.62x)&4,753(0.58x)&5,843(0.57x)&7,375(0.65x)&9,081(0.80x)\\
  \midrule
  Average &0.78x&0.78x&0.78x&0.81x&0.87x\\
     \bottomrule
      \end{tabular}   
    \end{table*}

\end{document}